\documentclass{aa}
\newif\iffigure
\figurefalse
\figuretrue

\DeclareRobustCommand{\erase}{\bgroup\markoverwith{\textcolor{red}{\rule[.5ex]{2pt}{2pt}}}\ULon}

\usepackage[switch]{lineno}
\usepackage{graphicx,natbib,url,twoopt}
\usepackage[varg]{txfonts}           %% A&A font choice
\usepackage{hyperref}              %% for pdflatex
\usepackage{pdfcomment}              %% for popup acronym meanings
\usepackage{acronym}                 %% for popup acronym meanings
\usepackage{ulem}                    % 取り消し線を引く
\usepackage{cases}
\usepackage{empheq}                  %Figs. 1,2 
\usepackage{here}
\usepackage{bm}
\usepackage{tabularx}
\usepackage[version=3]{mhchem}
\hypersetup{% hyperref option
 colorlinks=true,%
 linkcolor=blue,
 citecolor=blue,
}

\usepackage[
  draft,
  authormarkup=none,
  commandnameprefix=ifneeded
]{changes}% finalをdraftにするとハイライト表示
\definechangesauthor[color=black]{R1}
\definechangesauthor[color=black]{R2}
\definechangesauthor[color=black]{R3}
\definechangesauthor[color=black]{R4} % revised points
\definechangesauthor[color=black]{R5} % 2nd rev
\definechangesauthor[color=black]{R6} % 3rd rev
\definechangesauthor[color=black]{P2} % mentioning paper2
\usepackage{framed}
\usepackage{mdframed}

\newcommand*\patchAmsMathEnvironmentForLineno[1]{
  \expandafter\let\csname old#1\expandafter\endcsname\csname #1\endcsname
  \expandafter\let\csname oldend#1\expandafter\endcsname\csname end#1\endcsname
  \renewenvironment{#1}
     {\linenomath\csname old#1\endcsname}
     {\csname oldend#1\endcsname\endlinenomath}}
\newcommand*\patchBothAmsMathEnvironmentsForLineno[1]{
  \patchAmsMathEnvironmentForLineno{#1}
  \patchAmsMathEnvironmentForLineno{#1*}}
\AtBeginDocument{
\patchBothAmsMathEnvironmentsForLineno{equation}
\patchBothAmsMathEnvironmentsForLineno{align}
\patchBothAmsMathEnvironmentsForLineno{flalign}
\patchBothAmsMathEnvironmentsForLineno{alignat}
\patchBothAmsMathEnvironmentsForLineno{gather}
\patchBothAmsMathEnvironmentsForLineno{multline}
}

\defcitealias{kuwahara2026multi2}{KL26c}
\bibpunct{(}{)}{;}{a}{}{,}    %% natbib cite format used by A&A and ApJ

\makeatletter
\newcommand{\bibnote}[2]{\global\@namedef{#1note}{#2}}
\newcommand{\biblink}[2]{\global\@namedef{#1link}{#2}}
\newcommand{\Tabref}[1]{Table~\ref{#1}}
\newcommand{\Equref}[1]{Eq.~\ref{#1}}

\makeatother

\makeatletter
 \newcommandtwoopt{\citeads}[3][][]{%
   \nonstopmode%              %% fix to not stop at error message in latex
   \href{http://adsabs.harvard.edu/abs/#3}%
        {\def\hyper@linkstart##1##2{}%
         \let\hyper@linkend\@empty\citealp[#1][#2]{#3}}%   %% Rutten, 2000
   \biblink{#3}{\href{http://adsabs.harvard.edu/abs/#3}{ADS}}%
   \errorstopmode}            %% fix to resume stopping at error messages 
 \newcommandtwoopt{\citepads}[3][][]{%
   \nonstopmode%              %% fix to not stop at error message in latex
   \href{http://adsabs.harvard.edu/abs/#3}%
        {\def\hyper@linkstart##1##2{}%
         \let\hyper@linkend\@empty\citep[#1][#2]{#3}}%     %% (Rutten 2000)
   \biblink{#3}{\href{http://adsabs.harvard.edu/abs/#3}{ADS}}%
   \errorstopmode}            %% fix to resume stopping at error messages
 \newcommandtwoopt{\citetads}[3][][]{%
   \nonstopmode%              %% fix to not stop at error message in latex
   \href{http://adsabs.harvard.edu/abs/#3}%
        {\def\hyper@linkstart##1##2{}%
         \let\hyper@linkend\@empty\citet[#1][#2]{#3}}%     %% Rutten (2000)
   \biblink{#3}{\href{http://adsabs.harvard.edu/abs/#3}{ADS}}%
   \errorstopmode}            %% fix to resume stopping at error messages 
 \newcommandtwoopt{\citeyearads}[3][][]{%
   \nonstopmode%              %% fix to not stop at error message in latex
   \href{http://adsabs.harvard.edu/abs/#3}%
        {\def\hyper@linkstart##1##2{}%
         \let\hyper@linkend\@empty\citeyear[#1][#2]{#3}}%  %% 2000
   \biblink{#3}{\href{http://adsabs.harvard.edu/abs/#3}{ADS}}%
   \errorstopmode}            %% fix to resume stopping at error messages 

\renewcommand{\@biblabel}[1]{[#1]} % これは番号の出力形式（任意）

\newacro{ADS}{Astrophysics Data System}
\newacro{NLTE}{non-local thermodynamic equilibrium}
\newacro{NASA}{National Aeronautics and Space Administration}

\begin{document}
%\linenumbers
\authorrunning{Kuwahara and Lambrechts}
\titlerunning{Dust transport in envelopes of disk-embedded planets I.}
   % Title 
   % Gas-driven dust transport in envelopes of disk-embedded planets:
   % Gas–dust coupling in envelopes of disk-embedded planets:
   % How gas dynamics shape dust distribution in envelopes around disk-embedded planets:
   \title{Dust transport in envelopes of disk-embedded planets:}
   \subtitle{I. Convectively stable envelopes}
      \author{Ayumu Kuwahara\inst{1} 
          \thanks{\email{ayumu.kuwahara@sund.ku.dk}} 
          \and Michiel Lambrechts\inst{1}}

   \institute{Center for Star and Planet Formation, Globe Institute, University of Copenhagen, Øster Voldgade 5-7, 1350 Copenhagen, Denmark
         }

   \date{Received September XXX; accepted YYY}

  \abstract
    {
    \added[id=R2]{Planets embedded in protoplanetary disks accrete solids through their \added[id=R4]{gaseous envelopes}. The spatial distribution of these dust particles inside the envelopes of disk-embedded planets is poorly known.}
    We present high-resolution two- and three-dimensional multifluid simulations that follow the dynamics of gas and dust around \added[id=R2]{planets similar in mass to Earth.}
    \added[id=R4]{Our simulations resolve an outer recycling flow and an inner convectively stable envelope that is shielded from the recycling flow.}
    We identify strongly dust-depleted envelopes: the dust-to-gas ratio decreases radially inward and is reduced by more than two to four orders of magnitude in the deep interior ($<0.1\,R_{\rm B}$; Bondi radius) compared to its value at the outer edge of the envelope.
    Small grains\added[id=R2]{, with a dimensionless stopping time ${\rm St}\lesssim10^{-3}$,} remain entrained in the recycling flow and do not enter the envelope, whereas large grains (${\rm St}\gtrsim10^{-2}$) penetrate the envelope but settle rapidly onto the core \added[id=R2]{along the midplane}.
    \added[id=R4]{The resulting dust depletion in convectively-stable envelopes implies a substantial reduction in dust opacity throughout much of the envelope, facilitating cooling and a more rapid transition to runaway gas accretion.}
    \added[id=R2]{These results further suggest that enriching the \added[id=R4]{deep} envelope ($<0.1\,R_{\rm B}$) with dust or volatiles requires their delivery through large pebbles that then subsequently disintegrate, or sublimate, from their host grains in the deep envelope interior.}
    }
    

   \maketitle

%---------------------------------------------------------
%---------------------------------------------------------
%---------------------------------------------------------
%%%%%%%%%%%%%%%%%%%%%%%%%%%%%%%%%%%%%%%%%%%%%%%%%%%%%%%%%%
%%%%%%%%%%%%%%%%%%%%%%%%%%%%%%%%%%%%%%%%%%%%%%%%%%%%%%%%%%
%%%%%%%%%%%%%%%%%%%%%%%%%%%%%%%%%%%%%%%%%%%%%%%%%%%%%%%%%%

\section{Introduction}\label{sec:Introduction}

%---------------------------------------------------------
Dust particles of approximately millimeter–centimeter size are crucial for the growth of the \added[id=R1]{low-mass protoplanet ($\lesssim10\,M_\oplus$; Earth masses)} and the thermal evolution of \added[id=R1]{its} envelope, as they efficiently enter the envelope, settle toward the center, and accrete onto the planet \citep{Ormel:2010,Lambrechts:2012}. The envelope’s thermal evolution is regulated by the opacity, whose predominant source is dust particles at temperatures below approximately $1700$~K \citep{bell1994using}. The spatial distribution of these small grains therefore act as primary control parameters for the envelope’s cooling time and for the timing of runaway gas accretion \citep{Hori:2011,Lee:2015}.

However, the dust distribution, and thus the dust opacity\added[id=R2]{,} within envelopes remains poorly constrained.
\added[id=R1]{Previous studies have typically assumed a prescribed dust opacity, either a power-law expression or one taken from opacity tables, thereby bypassing the problem of determining the dust distribution within the envelope itself \citep{brouwers2020planets,zhu2021global}.}
In the outer envelope, the opacity is expected to inherit the background disk value, often assumed to be intersteller-medium (ISM)-like \citep[$\sim 1~\mathrm{cm^2\,g^{-1}}$; e.g.,][]{piso2014minimum}.
Deeper inside, dust growth can facilitate efficient settling, potentially reducing the opacity well below the ISM value \citep{ormel2014atmospheric,mordasini2014grain}.

It is therefore essential to quantify the dust distribution within the envelope \added[id=R1]{of low-mass \added[id=R2]{protoplanets}}.
The commonly adopted approximation of an unperturbed Keplerian flow around an embedded planet can substantially overestimate how efficiently dust enters the envelope \citep{Kuwahara:2020a,Kuwahara:2020b,okamura2021growth}. 
In reality, the planet’s gravity perturbs the disk gas and alters the flow topology \citep{Ormel:2013}. 
Recent three-dimensional (3D) hydrodynamical simulations identify a recycling layer in the outer envelope, in which disk gas enters at high latitude and exits near the midplane \citep[e.g.,][]{Ormel:2015b,Fung:2015,Kuwahara:2019}. 
Inside the envelope, depending on the temperature gradient, radiative and (or) convective layers develop \citep{rafikov2006atmospheres,Lambrechts:2017,Popovas:2018b,zhu2021global,KL26}. 
These gas dynamics affect dust dynamics and can regulate the dust mass flux, especially for tightly coupled grains.

A further understanding of the planet–envelope system requires treating the coevolution of gas and dust. 
\added[id=R1]{\added[id=R2]{Little work has been done} to characterize dust distribution in envelopes during dust accretion with hydrodynamical simulations.
\added[id=R2]{Recently,} \cite{krapp20223d} performed 3D global multifluid simulations of disk-planet interaction, finding an anisotropic dust distribution within the envelope.
\added[id=R3]{Local simulations offer a complementary approach by resolving the envelope at much higher spatial resolution, allowing us to fully explore the detailed interplay between envelope gas dynamics and dust transport.}
}

\added[id=R1]{Here we perform multifluid simulations of gas and dust in a local frame co-rotating with a planet to study dust transport within the envelope.} 
\added[id=R5]{In this first paper (Kuwahara \& Lambrechts 2026b; KL26b), we focus on the radiative end-member case in which the envelope remains nearly isothermal and convectively stable.}
\added[id=R4]{Such envelopes are likely to emerge in the outer parts of disks ($\gtrsim10$ au) where the envelope’s cooling time is short \citep[e.g.,][]{rafikov2006atmospheres,KL26}.}
\added[id=R5]{The companion paper investigates the opposite limit of a fully convective envelope \cite[][hereafter \citetalias{kuwahara2026multi2}]{kuwahara2026multi2}.
Together, these idealized models are intended to bracket the range of possible dust dynamics, rather than provide a fully self-consistent description of envelope thermodynamics.}
Because the gas flow structure depends on dimensionality\added[id=R1]{---for example, the recycling flows only appear in 3D---}we carry out simulations in both 2D and 3D.

The paper is organized as follows. 
Section~\ref{sec:Numerical method} describes the numerical setup for our 2D and 3D multifluid simulations. 
Section~\ref{sec:Numerical results} presents the emergence of a dust-depleted envelope in our simulations. 
In Section~\ref{sec:1D analytic model} we construct analytic formulae that reproduce the numerical results. 
\added[id=R2]{Sections~\ref{sec:Comparison to previous studies} and \ref{sec:Discussions} place our results in the context of previous work and discuss potential implications on envelope growth and composition.}
We summarize our findings in Section~\ref{sec:Conclusions}.

%---------------------------------------------------------
%---------------------------------------------------------
\section{Numerical methods}\label{sec:Numerical method}
We simulated gas and dust dynamics around a planet embedded in a non-self-gravitating disk with the Athena++ code with the multifluid dust module \citep{stone2020athena++,HuangBai2022}. Our simulations were performed in either 2D cylindrical or 3D spherical polar coordinates centered on a planet, where $r$ is the distance from the planet, $\theta$ the polar angle, and $\phi$ the azimuth angle. We used the default numerical settings of Athena++, such as the integration schemes, unless otherwise specified.

We assumed that the gas is a compressible, inviscid, and non-self-gravitating fluid, and the dust is a pressureless fluid. 
The Athena++ code solves the following sets of equations of gas and dust:
\begin{align}
    &\frac{\partial \rho_{\rm g}}{\partial t}+\nabla\cdot(\rho_{\rm g}\bm{v}_{\rm g})=0,\\
    &\frac{\partial (\rho_{\rm g}\bm{v}_{\rm g})}{\partial t}+\nabla\cdot(\rho_{\rm g} \bm{v}_{\rm g}\bm{v}_{\rm g})=-\nabla p+\rho_{\rm g}\bm{f}_{\rm src}\,,\\
    &\frac{\partial E}{\partial t}+\nabla\cdot\left[(E+p)\bm{v}_{\rm g}\right]=\rho_{\rm g}\bm{v}_{\rm g}\cdot\bm{f}_{\rm src}-\frac{e-e_{\rm 0}}{t_{\rm cool}},\label{eq:energy eq}\\
    &\frac{\partial \rho_{\rm d}}{\partial t}+\nabla\cdot(\rho_{\rm d}\bm{v}_{\rm d})=0,\\
    &\frac{\partial (\rho_{\rm d}\bm{v}_{\rm d})}{\partial t}+\nabla\cdot(\rho_{\rm d} \bm{v}_{\rm d}\bm{v}_{\rm d})=\rho_{\rm d}\bm{f}_{\rm src}+\rho_{\rm d}\frac{\bm{v}_{\rm g}-\bm{v}_{\rm d}}{t_{\rm s}}\label{eq:dust eom}.
\end{align}
Here $\rho$ is the density, $\bm{v}$ is the velocity, and $p$ is the pressure. The subscripts "g" and "d" denote gas and dust, respectively. In 2D, $\rho$ denotes surface density. \added[id=R1]{We continue to use $\rho$ to denote both the surface and volume densities, and use the superscripts "2D" and "3D" when explicitly specifying the dimensionality.}

We solved the energy equation including a thermal relaxation term on the right-hand side of \Equref{eq:energy eq}. 
The total energy density is given by $E=e+\rho_{\rm g} v_{\rm g}^2/2$ and the internal energy density by $e=p/(\gamma-1)$, with $\gamma=1.43$ and $e_0$ being the adiabatic index and its initial value.

\begin{figure}[tp]
    \centering
    \includegraphics[width=1\linewidth]{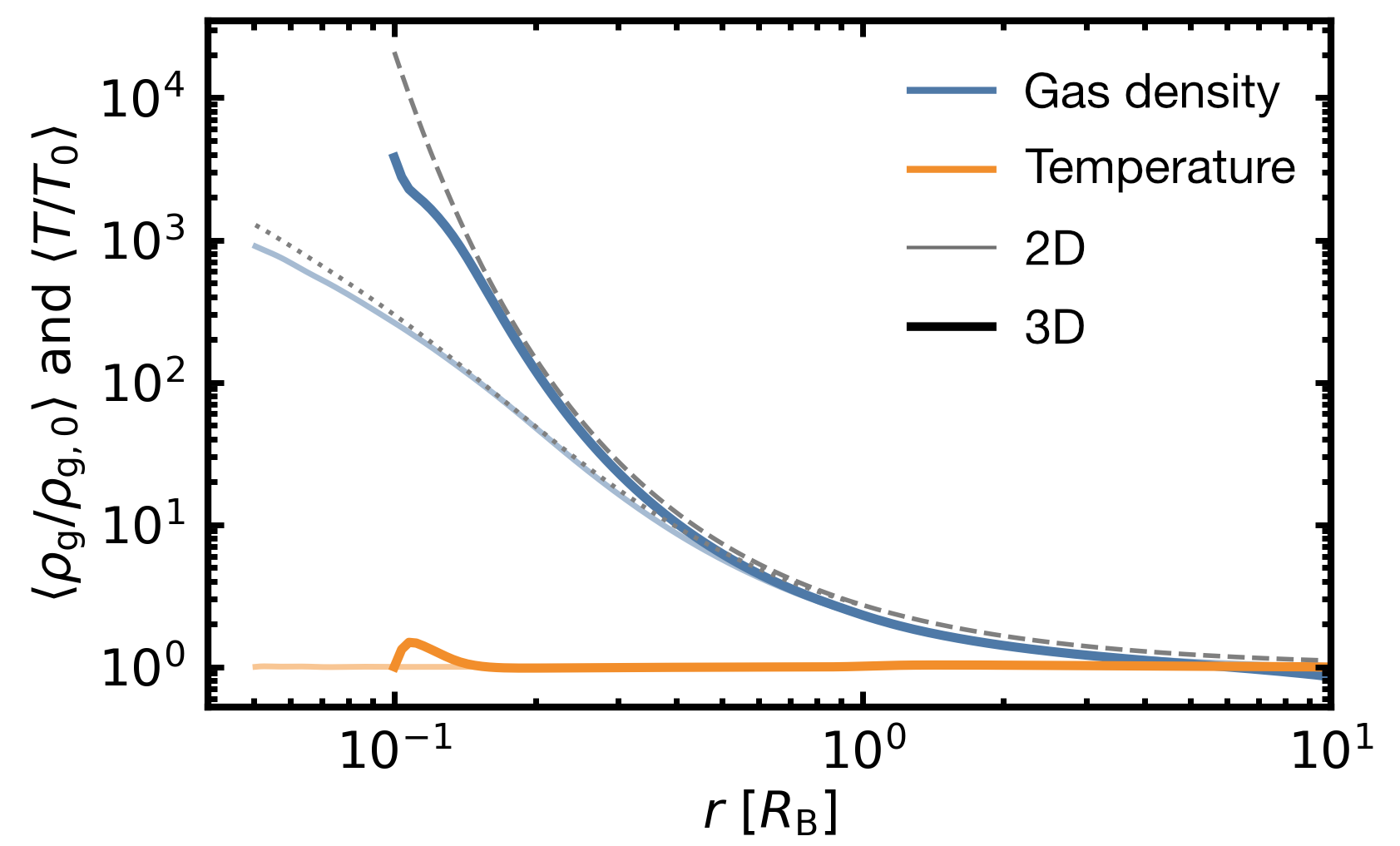}
    \caption{\added[id=R4]{Radial profiles of the gas density (blue) and temperature (orange). 
    The dashed curve assumes hydrostatic equilibrium (Eq.~\ref{eq:gas density model, 3D}), and the dotted curve is obtained by solving for the vortensity conservation together with radial force balance (Sect.~\ref{sec:1D analytic model}).}}
    \label{fig:1dslice_rho_and_temp}
\end{figure}

We parameterized the thermal relaxation timescale as the dimensionless quantity $\beta\equiv t_{\rm cool}\Omega$ \citep{Gammie:2001}. 
We set $\beta=1$ throughout the computational domain. 
\added[id=R5]{We prescribe the cooling time to construct an idealized envelope that remains nearly isothermal and convectively stable throughout the Bondi sphere \citep[Fig.~\ref{fig:1dslice_rho_and_temp};][]{KL26}.
This setup allows us to isolate dust dynamics in a convectively stable envelope without introducing uncertainties related to a self-consistent thermal structure (further discussed in Sect.~\ref{sec:Implications for thermal evolution of envelopes}).}
Appendix \ref{sec:Convergence tests} compares nearly-isothermal and isothermal runs, showing a quantitative agreement with each other. 
\added[id=R6]{A convectively stable layer is likely to develop when radiative cooling is sufficiently efficient, as determined by the opacity, accretion luminosity, and local disk conditions \citep{rafikov2006atmospheres,KL26}.
Appendices A and B of \citetalias{kuwahara2026multi2} present a 1D envelope model and cooling-time analysis.
These calculations suggest that nearly isothermal outer envelope layers can develop in low-opacity regions of the outer disk, although maintaining efficient cooling throughout the modeled envelope requires a relatively restricted range of opacity (e.g., $\kappa\simeq10^{-3}\,\mathrm{cm^2/g}$ at 10 au, Fig.\,B.1b in \citetalias{kuwahara2026multi2}).
We therefore regard the adopted setup as an idealized radiative end-member rather than as a generic outer-disk envelope structure.}
We neglected any heat sources: the accretion of solids, the latent heat from dust evaporation, and the radioactive heating, except the compressional heating due to the second term on the left hand side of \Equref{eq:energy eq}.
\added[id=R3]{We note that accretion heating, which can drive envelope convection, will be included in the second paper of this series \citepalias{kuwahara2026multi2}, where dust transport in convective envelopes will be investigated.
}

In a frame co-moving with a planet, the source terms include the gravity of the planet, the Coriolis and the tidal forces, $\bm{f}_{\rm src}=\bm{f}_{\rm grav}+\bm{f}_{\rm cor}+\bm{f}_{\rm tid}$, $\bm{f}_{\rm grav}=-\nabla\Phi_{\rm p}$, $\bm{f}_{\rm cor}=-2\bm{e}_z\times\bm{v}_{\rm g}$, and $\bm{f}_{\rm tid}=3x\bm{e}_x-z\bm{e}_z$. We implemented the gravitational potential of the planet as
\begin{empheq}[left = {\Phi_{\rm p}=\empheqlbrace \,}]{alignat = 2}
    &-\frac{GM_{\rm p}}{\sqrt{r^2+r_{\rm sm}^2}}\,f_{\rm inj}&&\quad{\text{(for the gas, 2D)}},\label{eq:plummer potential}\\
    &-\frac{GM_{\rm p}}{r}\,f_{\rm sm}\,f_{\rm inj}&&\quad{\text{(for the gas, 3D)}},\label{eq:force-free potential}\\
    &-\frac{GM_{\rm p}}{r}\, f_{\rm inj}&&\quad{\text{(for the dust)},}
\end{empheq}
where $G$ is the gravitational constant and $M_{\rm p}$ is the planet mass. 
\added[id=R4]{We did not apply gravitational smoothing to the dust, as it would artificially reduce the infall velocity and produce a numerical pile-up within the envelope.
By contrast, the gas potential is smoothed, as resolving the balance between planetary gravity and the steep pressure gradient force near the inner boundary is numerically challenging without smoothing \citep[e.g.,][]{Ormel:2015b}.
}
For the gas, we adopted the Plummer smoothing in 2D \citep{plummer1911problem}, and a force-free smoothing at the inner boundary in 3D where the gravitational potential is smoothed by \citep{Fung:2019,zhu2021global}:
\begin{align}
    f_{\rm sm}=\frac{(r-r_{\rm in})^2}{(r-r_{\rm in})^2+r_{\rm sm}^2}.
\end{align}
Here $r_{\rm in}$ is the size of the inner boundary and $r_{\rm sm}$ the smoothing length. We set $r_{\rm sm}=0.1\,R_{\rm B}$ in 2D and $r_{\rm sm}=0.1\,r_{\rm in}$ in 3D. 
For the fiducial runs, $r_{\rm in}$ is approximately 5 (10) times the physical size of the core in 2D (3D). 
The gravity of the planet is gradually inserted into the disk to prevent shock formation. 
Following \cite{Ormel:2015a}, we used the following injection function 
\begin{align}
    f_{\rm inj}=1-\exp\Bigg[-\frac{1}{2}\Bigg(\frac{t}{t_{\rm inj}}\Bigg)^2\Bigg],
\end{align}
where $t$ is the time and $t_{\rm inj}=5\,\Omega_0^{-1}$ is the injection time and $\Omega_0$ is the orbital frequency. 
Appendix \ref{sec:Convergence tests} explores the dependence on the gravitational potential formula.

\added[id=R4]{We computed the drag force on the dust component as
\begin{align}
    \bm{F}_{\rm drag}=-\frac{C_{\rm D}}{2}\pi s^2\rho_{\rm g}u\bm{u},
\end{align}
where $\bm{u}\equiv \bm{v}_{\rm d}-\bm{v}_{\rm g}$ and $u\equiv |\bm{u}|$. 
For $s<9l_{\rm mfp}/4$, we adopted the Epstein drag law,
\begin{align}
    C_{\rm D}=\frac{8c_{\rm s}}{3u},\label{eq:Epstein drag coefficient}
\end{align}
where the mean free path is $l_{\rm mfp}=\mu m_{\rm H}/(\rho_{\rm g}\sigma_{\rm mol})$ with $\mu=2.34$, $m_{\rm H}$ the proton mass, and $\sigma_{\rm mol}=2\times10^{-15}\,\mathrm{cm^2}$ \citep[e.g.,][]{Nakagawa:1986}.
For larger particles, $s\ge 9l_{\rm mfp}/4$, we used \citep{Weidenschilling:1977}
\begin{align}
    C_{\rm D}=
    \begin{cases}
        24\,\mathrm{Re}_{\rm p}^{-1} & (\mathrm{Re}_{\rm p}<1),\\
        24\,\mathrm{Re}_{\rm p}^{-0.6} & (1<\mathrm{Re}_{\rm p}<800),\\
        0.44 & (\mathrm{Re}_{\rm p}>800),\label{eq:nonlinear drag coefficient}
    \end{cases}
\end{align}
where $\mathrm{Re}_{\rm p}=2su/\nu$ is the particle Reynolds number with $\nu=l_{\rm mfp}c_{\rm s}/2$ being the viscosity.
The stopping time is defined as
\begin{align}
    t_{\rm s}\equiv \frac{m_{\rm p}u}{|\bm{F}_{\rm drag}|},
    \label{eq:stopping time}
\end{align}
where $m_{\rm p}=4\pi s^3\rho_\bullet/3$ is the particle mass and $\rho_\bullet=3\,\mathrm{g\,cm^{-3}}$ is the internal density of the dust.
Gas drag acceleration is then applied via the second term on the right-hand side of Eq.~\ref{eq:dust eom}.
Dust diffusion and backreaction on the gas were neglected.
}

\begin{figure}[tp]
    \centering
    \includegraphics[width=1\linewidth]{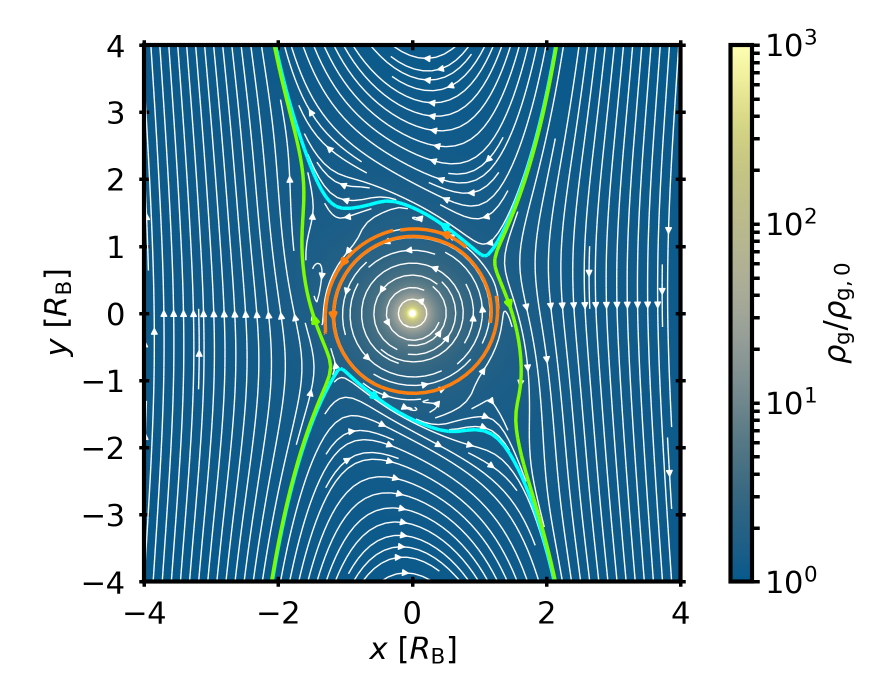}
    \caption{Gas flow field around an embedded planet in the 2D run with streamlines, showing the three distinct regions: the shear, the horseshoe, and the envelope. \added[id=R1]{The orange, cyan, and green curves mark the outer envelope, outer horseshoe and inner shear streamlines, respectively.} The background color shows the gas surface density.}
    \label{fig:slice_xy_rhog_2d}
\end{figure}

%-----------------------------------
%-----------------------------------
\subsection{Code units, simulation parameters, and initial condition}\label{sec:Code units, simulation parameters, and initial condition}

\begin{table*}[htbp]
\caption{Parameters of hydrodynamical simulations.}
%\centering
\resizebox{\textwidth}{!}{
\begin{tabular}{lccccccccc}\hline\hline
     & $m$ & St or $s$ & $\beta$ & Resolution & $t_{\rm end}$ [$\Omega_0^{-1}$] & $r_{\rm in}$ [\added[id=R4]{$R_{\rm B}$}] & $r_{\rm out}$ [\added[id=R4]{$R_{\rm B}$}] & Include $(\bm{f}_{\rm tid})_z$ & $\Phi_{\rm p}$ \\ \hline
     Fiducial runs (2D)  & $0.1$ & $10^{-3},\,10^{-2},\,10^{-1}$ & 1 & $(N_r,\,N_\phi)=(256,\,256)$ & 100 & $0.05$ & $10$ & - & \Equref{eq:plummer potential} \\
     Fiducial runs (3D)  & $0.1$ & $10^{-3},\,10^{-2},\,10^{-1}$ & 1 & $(N_r,\,N_\theta,\,N_\phi)=(128,\,32,\,128)$ & 100 & $0.1$ & $10$ & yes & \Equref{eq:force-free potential}\\\hline
     Fixed dust size runs (2D) & $0.1$ & $^\ast$0.01 cm, 0.1 cm, 1 cm, 10 cm & 1 & $(N_r,\,N_\phi)=(256,\,256)$ & 100 & $0.05$ & $10$ & - &  \Equref{eq:plummer potential} \\\hline
     Convergence tests   & $0.1$ & $10^{-2}$ & Isothermal & $(N_r,\,N_\phi)=(256,\,256)$ & 100 & 0.05 & $10$ & -  & \Equref{eq:plummer potential} or \Equref{eq:force-free potential}\\
                         & $0.1$ & $10^{-2}$ & 1 & $(N_r,\,N_\phi)=(512,\,512)$ & 100 & 0.05 & $10$ & - & \Equref{eq:plummer potential}\\
                         & $0.1$ & $10^{-2}$ & Isothermal & $(N_r,\,N_\theta,\,N_\phi)=(128,\,32,\,128)$ & 100 & 0.1 & $10$ & yes & \Equref{eq:plummer potential} or \Equref{eq:force-free potential}\\
                         & $0.1$ & $10^{-2}$ & 1 & $(N_r,\,N_\theta,\,N_\phi)=(128,\,32,\,128)$ & 50 & $0.1$ & $10$ & no & \Equref{eq:force-free potential}\\
                         & $0.1$ & $10^{-2}$ & 1 & $(N_r,\,N_\theta,\,N_\phi)=(128,\,32,\,128)$ & 50 & $0.05$ & $10$ & yes & \Equref{eq:force-free potential}\\
                         & $0.1$ & $10^{-2}$ & 1 & $(N_r,\,N_\theta,\,N_\phi)=(256,\,64,\,256)$ & 10 & $0.1$ & $10$ & yes  & \Equref{eq:force-free potential}\\\hline
\end{tabular}
}
\tablefoot{The following columns give the dimensionless thermal mass, the Stokes number or the dust size, the dimensionless cooling time, the resolution, the calculation time, the size of the inner boundary, the size of the outer boundary, and the gravitational potential formula. \added[id=R2]{The run with the asterisk ($s=0.01$ cm) does not reach the steady state within the calculation time (Appendix~\ref{sec:Limitations for simulations with very small dust grains}).}}
\label{tab:hydro simulations}
\end{table*}

\begin{figure*}[tp]
    \centering
    \includegraphics[width=1\linewidth]{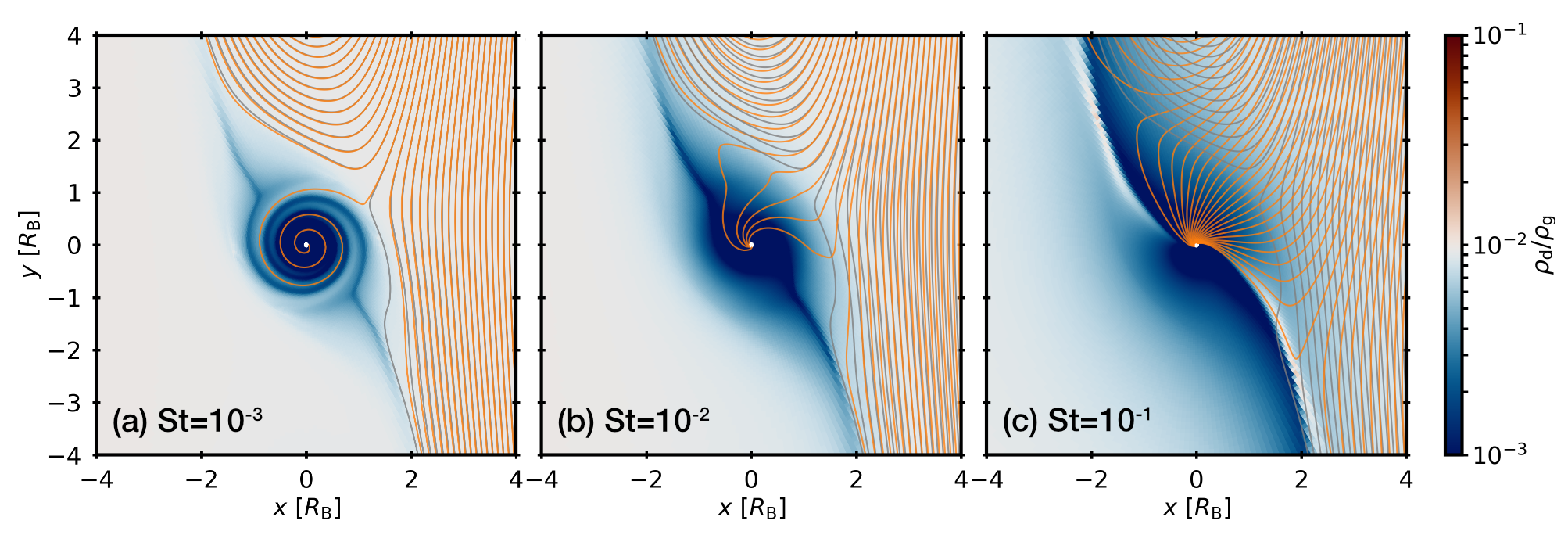}
    \caption{\added[id=R1]{Reduction of the dust-to-gas ratio towards a planetary core in the 2D runs, for particles with different Stokes numbers. Streamlines in the midplane of gas (gray) and dust (orange) originating from the first quadrant.} The background color shows the column dust-to-gas ratio, \added[id=R1]{$Z=\rho_{\rm d}^{\rm 2D}/\rho_{\rm g}^{\rm 2D}$}. We note that the color bar is saturated below $10^{-3}$. }
    \label{fig:slice_xy_d_to_g_2d}
\end{figure*}

Our simulations were performed in the units of $H_{\rm g,0}=c_{\rm s,0}=\Omega_0=\rho_{\rm g,0}^{\rm 2D}=\rho_{\rm g,0}^{\rm 3D}=1$, where $H_{\rm g,0}$ is the gas scale height, \added[id=R1]{$c_{\rm s,0}$} the isothermal sound speed, $\rho_{\rm g,0}^{\rm 2D}$ the initial gas surface density, and $\rho_{\rm g,0}^{\rm 3D}$ the midplane gas density at the planet orbital location, $a_{\rm p}$. 
\added[id=R1]{The envelope is nearly isothermal, so that $c_{\rm s}\simeq c_{\rm s,0}$ holds throughout the computational domain.} 
Since we neglect the self-gravity of the disk gas, we can introduce another normalization for the planetary mass,
\begin{align}
    m\equiv\frac{R_{\rm B}}{H_{\rm g,0}}&=\frac{M_{\rm p}}{M_{\rm th}}\simeq0.11\,\Bigg(\frac{M_{\rm p}}{M_\oplus}\Bigg)\Bigg(\frac{M_\ast}{M_\odot}\Bigg)^{-1}\Bigg(\frac{0.03}{h}\Bigg)^3.
\end{align}
Here, $R_{\rm B}=GM_{\rm p}/c_{\rm s,0}^2$ is the Bondi radius, $M_{\rm th}=M_\ast h^3$ the thermal mass, $M_\ast$ the stellar mass, $h$ the disk aspect ratio, and $M_\odot$ the solar mass. We set $m=0.1$ in this study. The Hill radius in code units is given by $R_{\rm H}/$\added[id=R4]{$H_{\rm g,0}$}$=(m/3)^{1/3}\simeq0.32$.
\added[id=R5]{We confirmed that the modeled envelope mass remains much smaller than the planet mass, $M_{\rm env}/M_{\rm p}\lesssim0.01$.}
\added[id=R6]{This estimate includes only the gas at $r\geq r_{\rm in}$, leaving the mass and self-gravity of the unresolved deeper envelope unconstrained.}

The dimensionless stopping time of the dust, referred to as the Stokes number is defined by
\begin{align}
    {\rm St}=t_{\rm s}\Omega_0.
\end{align}
In fiducial runs, we assumed a fixed Stokes number throughout the computational domain, ${\rm St}=10^{-3},\,10^{-2},$ and $10^{-1}$. 
We also performed fixed-size runs with constant dust size, $s=0.01$ cm, 0.1 cm, 1 cm, and 10 cm. 
Dust physics such as growth, fragmentation, erosion, and ablation, were not included, which will be discussed in Sect.~\ref{sec:Dust processing within envelopes}. 
We note that in the case of 0.01 cm-sized dust, the simulation does not reach the steady state within the calculation time (Appendix~\ref{sec:Limitations for simulations with very small dust grains}).

We assumed a vertically stratified density profile for the initial condition, 
\begin{align}
    \rho_{i}=\rho_{i,0}\exp\Bigg[-\frac{1}{2}\bigg(\frac{z}{H_{i,0}}\bigg)^2\Bigg],\label{eq:density profile}
\end{align}
Here $i$ corresponds to "g" or "d", $\rho_{i,0}$ is the initial density at the planet location, and $H_{i,0}$ is the scale height.
The floor values of the density were set to $10^{-6}$ and $10^{-8}$ for the gas and dust, respectively. 
We defined the column dust-to-gas ratio,
\begin{align}
    Z\equiv\frac{\rho_{\rm d}^{\rm 2D}}{\rho_{\rm g}^{\rm 2D}},
\end{align}
and $Z_0\equiv\rho_{\rm d,0}^{\rm 2D}/\rho_{\rm g,0}^{\rm 2D}=0.01$ being its initial value. 
The dust-to-gas ratio is defined by
\begin{align}
    \epsilon\equiv\frac{\rho_{\rm d}^{\rm 3D}}{\rho_{\rm g}^{\rm 3D}},
\end{align}
with $\epsilon_0\equiv\rho_{\rm d,0}^{\rm 3D}/\rho_{\rm g,0}^{\rm 3D}=0.01$ being its initial value. 
We note that, in the case of 3D, the column dust-to-gas ratio differs from that in 2D. 
Although we do not include a turbulence stirring, we prescribe the dust scale height in the initial condition. 
We set $H_{\rm d,0}=0.1\,H_{\rm g,0}=R_{\rm B}$, so that the envelope is initially filled with dust.

\added[id=R4]{A Keplerian shear flow was applied as an initial background velocity field, neglecting the headwind of the gas due to a global pressure gradient in a disk. 
Appendix~\ref{sec:Additional simulations} shows that including a headwind has only a minor impact on the dust-to-gas ratio within the envelope. Neglecting the headwind, however, leads to an overestimate of the dust accretion rate onto the planet \citep{Liu:2018}, implying that resulting dust-to-gas ratios should be regarded as upper limits.}
The initial velocities of gas and dust were 
\begin{align}
    \frac{\bm{v}_{\rm g,\infty}}{c_{\rm s,0}}=\frac{\bm{v}_{\rm d,\infty}}{c_{\rm s,0}}=-\frac{3}{2}\frac{x}{H_{\rm g,0}}\,\bm{e}_y.
\end{align}
\added[id=R2]{The parameters of our simulations are summarized in Table~\ref{tab:hydro simulations}.}

%---------------------------------------------------------
%---------------------------------------------------------
\subsection{\added[id=R1]{Resolutions and boundary conditions}}

We used a log-spaced grid in the radial coordinate ranging from $r_{\rm in}$ to $r_{\rm out}$, whereas the polar and azimuth angles are uniformly divided. The size of the inner boundary was set to $0.05\,R_{\rm B}$ ($0.1\,R_{\rm B}$) in the 2D (3D) fiducial runs, respectively. 
These values correspond to approximately 5 and 10 times the physical radius of the planet, which is given \added[id=R4]{by \citep{Kuwahara:2020a}
\begin{align}
    \frac{R_{\rm p}}{H_{\rm g,0}}\simeq1.4\times10^{-3}\,\bigg(\frac{m}{0.1}\bigg)^{1/3}\bigg(\frac{a_{\rm p}}{\text{1 au}}\bigg)^{-1}.\label{eq:core radius}
\end{align}
T}hus, our computational domain covers a wide range of the envelope. 
The numerical resolution is given in \Tabref{tab:hydro simulations}. \added[id=R1]{In the 2D (3D) fiducial runs, the Bondi radius is resolved by 145 (64) grids in the radial direction. Appendix \ref{sec:Convergence tests} provides the resolution test.}

For the radial direction we set a reflecting and outflow boundary condition at the inner boundary for the gas and dust, respectively. 
When a heating source is absent, unphysical energy flux may occur at the inner boundary, which can be caused by the reflective boundary condition. 
To prevent this, following \cite{Kurokawa:2018} we set an instantaneous cooling at the inner boundary, $\beta=10^{-4}$. 
At the outer boundary, we fixed the density and the velocity to the initial values \added[id=R4]{for both gas and dust, thereby maintaining a constant mass flux within the local domain.}
In 3D runs, we only considered the upper half region of the disk, $\theta\in[0,\,\pi/2]$. 
We used a reflecting condition at the midplane, $\theta=\pi/2$. 
On the pole we used the polar boundary condition, in which the physical quantities in the ghost cells are copied from the other side of the pole \citep{stone2020athena++}. 
We considered the full range of the azimuthal angle, $\phi\in[0,\,2\pi]$.

%-----------------------------------
%-----------------------------------

%---------------------------------------------------------
%---------------------------------------------------------

%---------------------------------------------------------
%---------------------------------------------------------
\section{Numerical results}\label{sec:Numerical results}
We find that the dust-to-gas density ratio within the envelope decreases radially inward by several orders of magnitude. This conclusion holds in both 2D and 3D. We begin by summarizing the 2D gas and dust dynamics (Sect. \ref{sec:Dynamics of gas and dust in 2D}).  Subsequent sections describe the corresponding 3D behavior (Sect. \ref{sec:Dynamics of gas and dust in 3D}) and \added[id=R1]{the difference between assuming constant Stokes and constant particle radius (Sect.~\ref{sec:Simulations with fixed-size particles}). We} introduce a 1D analytic model that reproduces the numerical results (Sect. \ref{sec:1D analytic model}).

\begin{figure}[tp]
    \centering
    \includegraphics[width=1\linewidth]{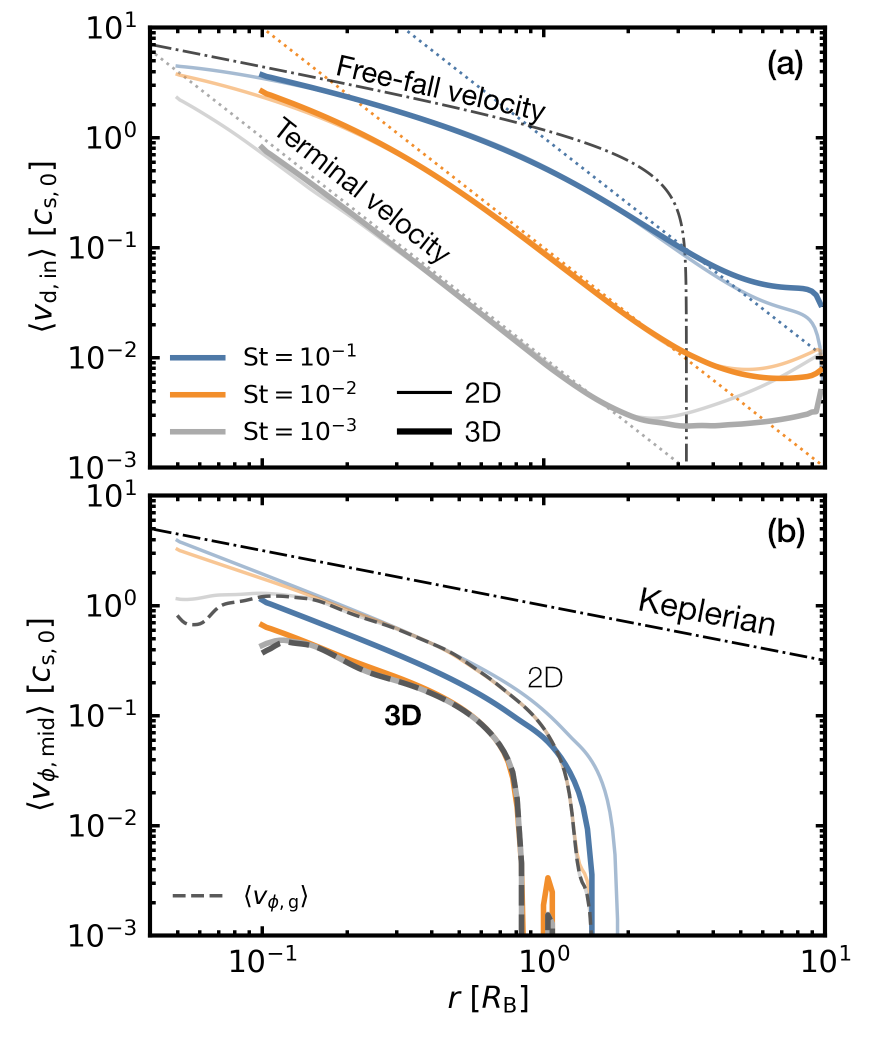}
    \caption{\added[id=R4]{Radial infall velocity of dust (\textit{top}) and azimuthal velocity of gas and dust at the midplane (\textit{bottom}). 
    The thin curves show azimuthally averaged results from 2D runs, while thick curves show shell-averaged (panel a) and azimuthally-averaged (panel b) results from 3D runs. 
    \textit{Top:} The dotted curves show the terminal velocity for different Stokes numbers, and the dot-dashed curve shows the free-fall velocity. 
    \textit{Bottom:} The dashed curves show the gas velocity, and the dot-dashed curve shows the Keplerian velocity.}}
    \label{fig:1dslice_vr_vphi}
\end{figure}

\begin{figure}[tp]
    \centering
    \includegraphics[width=1\linewidth]{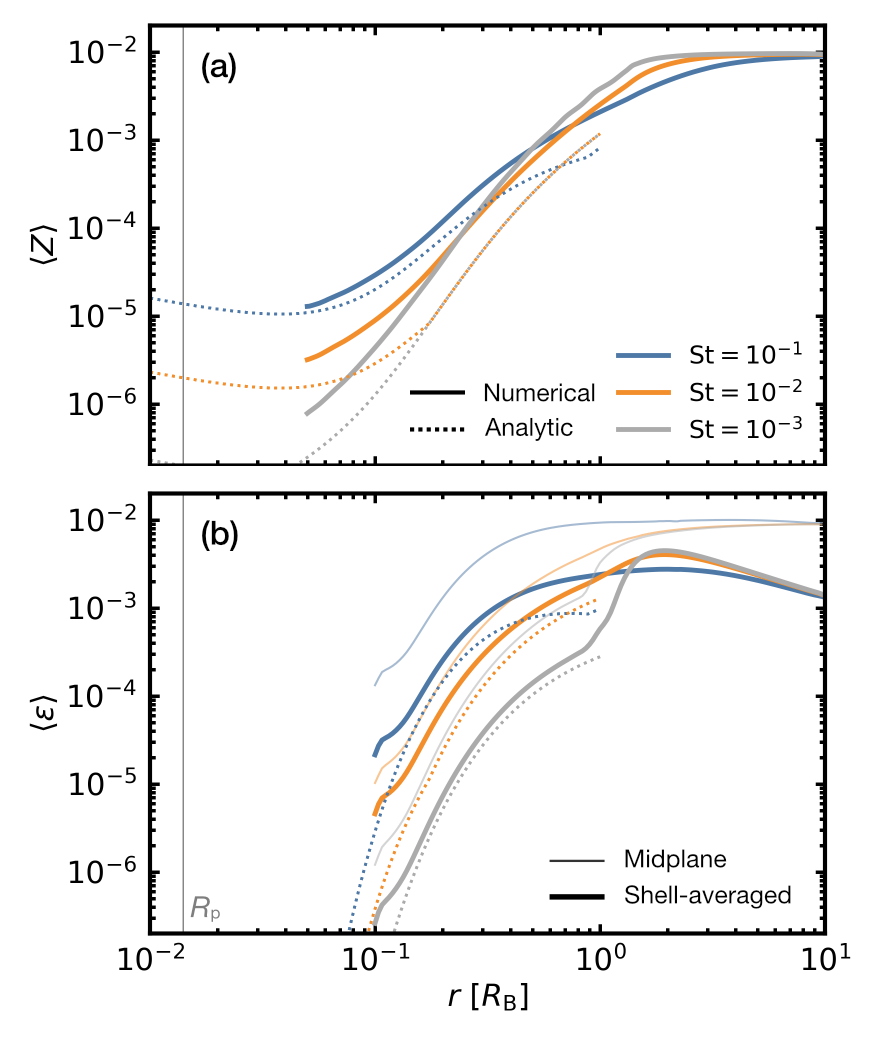}
    \caption{Dust-to-gas ratio for different Stokes numbers. \textit{Top}: azimuthally averaged value obtained from the 2D runs. \textit{Bottom}: Results from the 3D runs. The thick and thin solid curves correspond to the shell averagd value and the azimuthally averaged value at the midplane, respectively. The dotted curve is a 1D model introduced in Sect. \ref{sec:1D analytic model}, which is full analytic in the 3D, but semi-analytic in the 2D \added[id=R1]{case}. \added[id=R4]{The vertical line marks the core radius (Eq.~\ref{eq:core radius}).}}
    \label{fig:1dslice_d_to_g}
\end{figure}

%---------------------------------------------------------
%---------------------------------------------------------
\subsection{Dynamics of gas and dust in 2D}\label{sec:Dynamics of gas and dust in 2D}
Planets embedded in disks perturb the surrounding gas, thereby affecting dust dynamics. Because the gas flow past an embedded planet has been extensively studied \added[id=R1]{in both 2D and 3D \citep[e.g.,][]{Ormel:2015a,Ormel:2015b,Fung:2015}}, we highlight only the features that are key to understanding our results. The gas flow field separates into three characteristic regions: Keplerian shear, horseshoe, and the envelope. Figure \ref{fig:slice_xy_rhog_2d} shows the surface density and streamlines of the gas in the $x$-$y$ midplane. The Keplerian shear flow extends for $|x|\gtrsim R_{\rm B}$. The horseshoe flow exists in the upstream–downstream region along the planet’s orbit. An isolated inner envelope forms approximately within the Bondi radius and rotates prograde due to the Coriolis force. The gas surface density increases toward the planet by more than three orders of magnitude in the innermost region. 

Dust dynamics are inherited from gas dynamics and therefore depend on the Stokes number. For ${\rm St}=10^{-3}$, the dust is tightly coupled to the gas. The dust streamlines closely follow those of the gas (Fig. \ref{fig:slice_xy_d_to_g_2d}a; orange and gray solid curves). The dust coming from the narrow band between the horseshoe and shear regions enters the envelope, circulates prograde, and eventually accretes onto the planet. Deviations between gas and dust streamlines grow with increasing ${\rm St}$ (Fig. \ref{fig:slice_xy_d_to_g_2d}b and c). For ${\rm St}=10^{-1}$, dust is accreted onto the planet from a wide range of impact parameters.

\begin{figure*}[tp]
    \centering
    \includegraphics[width=1\linewidth]{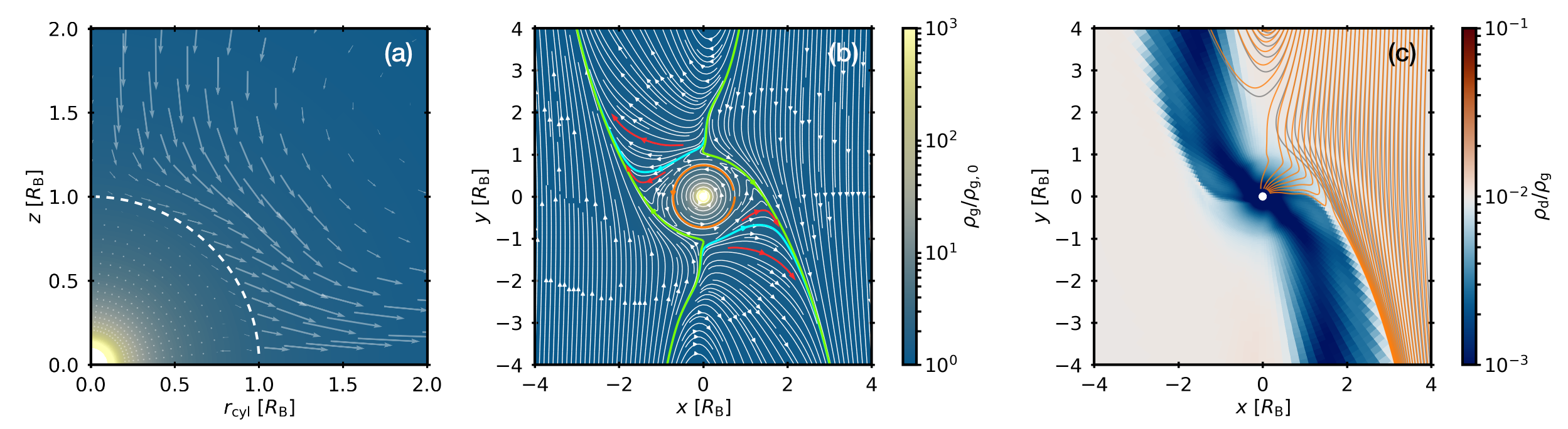}
    \caption{Flow field of gas and dust around an embedded planet in the 3D run. \textit{Left}: \added[id=R5]{Gas density and velocity vector of the gas, averaged over the azimuth $\phi\in[-\pi/2,\,\pi/2]$.} \textit{Middle}: Midplane slice with gas streamlines. \added[id=R1]{The orange, cyan, and green curves mark the outer envelope, outer horseshoe and inner shear streamlines, respectively. We note that many streamlines in the horseshoe region originate at high latitude---a genuine 3D recycling flow that cannot be fully captured in a midplane projection (red arrows).} \textit{Right}: \added[id=R1]{Streamlines in the midplane of gas (gray) and dust (orange) originating from the first quadrant.} The background color shows the dust-to-gas ratio. We set ${\rm St}=10^{-2}$.}
    \label{fig:slice_rhog_d_to_g_3d}
\end{figure*}

\begin{figure*}[tp]
    \centering
    \includegraphics[width=1\linewidth]{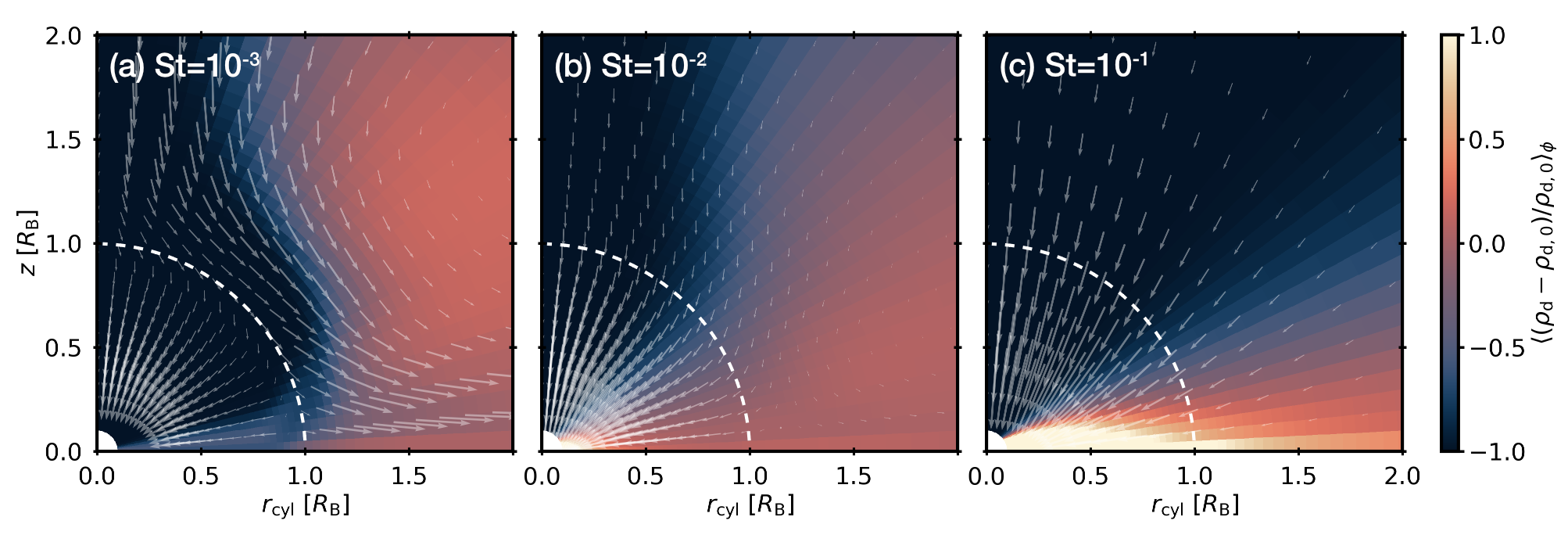}
    \caption{\added[id=R5]{Dust density and velocity vector of the dust, averaged over the azimuth $\phi\in[-\pi/2,\,\pi/2]$, for various Stokes numbers (${\rm St}=10^{-3},\,10^{-2}$, and $10^{-1}$).} The background color shows the deviation from the initial dust distribution. \added[id=R1]{The brighter and darker regions correspond to, respectively, regions where dust accumulates and depletes.}}
    \label{fig:slice_xz_rhog_rhod_azimuthal_average}
\end{figure*}

The envelope is dust-depleted, which is shown by the color contour in Figure~\ref{fig:slice_xy_d_to_g_2d}. 
Small dust grains are strongly coupled to the surrounding gas flow. 
Therefore the envelope is effectively shielded from the incoming dust flux, and only the dust grains initially resided within the envelope slowly settle toward the planet. 
Large dust grains can penetrate into the envelope but rapidly sediment. 
Consequently, starting from an initial value of 0.01, the column dust-to-gas ratio drops by orders of magnitude inside the envelope. 

The dust infall velocity is constrained by the smaller of the terminal and free-fall velocities\added[id=R4]{, 
\begin{align}
    &v_{\rm d,in}(r)=\min(v_{\rm term},\,v_{\rm ff}),\label{eq:v_d,in}\\
    &v_{\rm term}({\rm St},r)=\frac{m\,{\rm St}}{(r/H_{\rm g,0})^2}\,c_{\rm s,0},\\
    &v_{\rm ff}(r)=\sqrt{\frac{2m}{r/H_{\rm g,0}}\Bigg(1-\frac{r}{R_{\rm H}}\Bigg)}\,c_{\rm s,0},
    \label{eq:v ff}
\end{align}in} code units. Here we define the free-fall speed as the drag-free velocity obtained from energy conservation between $R_{\rm H}$ and $r$, assuming that the dust enters the Hill sphere with negligible kinetic energy.
Figure \ref{fig:1dslice_vr_vphi}a compares these expressions (dotted and dot-dashed curves) with the numerically obtained values (solid curves). For dust tightly coupled to the gas, ${\rm St}=10^{-3}$, $v_{\rm d,in}(r)$ follows the terminal speed closely. For dust marginally coupled to the gas (${\rm St}\gtrsim10^{-2}$), $v_{\rm d,in}(r)$ transitions to the free-fall speed in the deep envelope, $r\lesssim0.1\, R_{\rm B}$. 
\added[id=R4]{We note that the expression for $v_{\rm term}$ assumes linear drag, for which the stopping time is independent of the relative velocity. 
In the nonlinear regime, this assumption breaks down. 
We revisit this point in Sects.~\ref{sec:Simulations with fixed-size particles} and \ref{sec:1D analytic model}.}

\added[id=R3]{The azimuthal velocity of both gas and dust is sub-Keplerian throughout the computational domain (Fig.~\ref{fig:1dslice_vr_vphi}b). 
At $r=0.1\,R_{\rm B}$, the gas azimuthal velocity, $v_{\phi,{\rm g}}$, is reduced by a factor of approximately 3 relative to the Keplerian velocity in the 2D runs, and by a factor of approximately $10$ in the 3D runs, although gas may rotate closer to Keplerian velocities at even closer distances to the core that are not resolved here.
The dust azimuthal velocity closely follows $v_{\phi,{\rm g}}$, and remains smaller than the infall velocity throughout the envelope. This indicates that, at least between down to $0.1\,R_{\rm B}$, dust transport is dominated by infall rather than rotation.
}

\begin{figure}[tp]
    \centering
    \includegraphics[width=\linewidth]{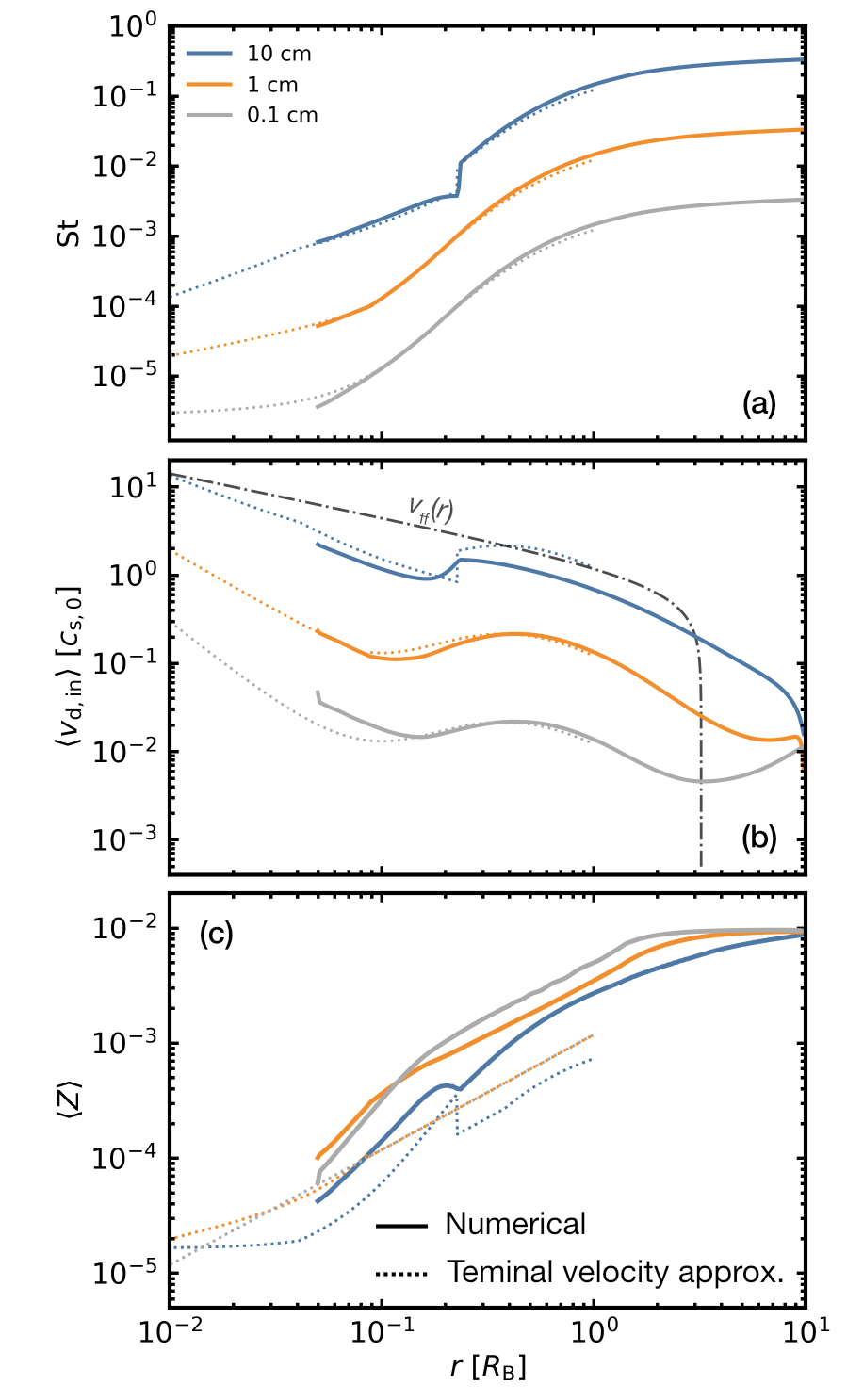}
    \caption{Stokes number, infall velocity of dust, and column dust-to-gas ratio obtained from the 2D, fixed dust size runs. All solid curves show the azimuthally averaged values. \added[id=R4]{The dotted curves are computed under the terminal-velocity approximation and are truncated at $r=R_{\rm B}$, since the gas density is evaluated only within the Bondi radius (Eqs.~\ref{eq:vortensity conservation} and \ref{eq:radial force balance}; Sect.~\ref{sec:1D analytic model}).}
    }
    \label{fig:1dslice_fixed_size}
\end{figure}

Figure \ref{fig:1dslice_d_to_g}a shows the azimuthally averaged column dust-to-gas ratio \added[id=R1]{as a function of radius} for different Stokes numbers, which decreases by 3--4 orders of magnitude \added[id=R1]{in the deep envelope relative to the value at the outer edge.} 
A 1D analytic model shown by the dotted curve in Fig. \ref{fig:1dslice_d_to_g}a is provided in Sect.~\ref{sec:1D analytic model}, where we demonstrate that the St-dependence of the dust-to-gas ratio is determined by \added[id=R4]{the dust accretion rate onto the planet (the envelope-penetration rate)} and the dust infall velocity.
We next assess whether this trend persists in 3D.

%---------------------------------------------------------
%---------------------------------------------------------
\subsection{Dynamics of gas and dust in 3D}\label{sec:Dynamics of gas and dust in 3D}
We find that the envelope remains dust-depleted in 3D, even though vertical gas flows modify dust motion. In 3D, the envelope interacts dynamically with the surrounding disk through a process known as recycling. Figures \ref{fig:slice_rhog_d_to_g_3d}a and b show the vertical and midplane slices of the gas flow field. The recycling flow is characterized by polar inflow and midplane outflow (Fig. \ref{fig:slice_rhog_d_to_g_3d}a). Despite the complexity of recycling flow, a gas inside approximately $\lesssim R_{\rm B}$ remains bound to a planet, because a positive entropy gradient suppresses the inflow penetrating the envelope. Owing to the isolation of the inner envelope, gas streamlines at the midplane resemble their 2D counterparts (Figs. \ref{fig:slice_xy_rhog_2d} and \ref{fig:slice_rhog_d_to_g_3d}b). 

The midplane outflow of the gas, \added[id=R1]{originating from high altitudes,} blows dust away from the planet, producing a low dust-to-gas density ratio in the upper left and lower right corners in Fig. \ref{fig:slice_rhog_d_to_g_3d}c. This feature is not observed in the 2D runs, where the midplane outflow is absent, and thus represents a characteristic unique to the 3D runs \citep{Kuwahara:2020a}.

Vertical dust motion is governed by settling and the polar gas inflow. Since the polar inflow does not penetrate the inner envelope, dust that tightly coupled to the gas is carried toward the midplane without entering the envelope. As in the 2D runs, the envelope is shielded from the incoming dust flow. This leads to dust depletion throughout the Bondi sphere (Fig. \ref{fig:slice_xz_rhog_rhod_azimuthal_average}a, ${\rm St}=10^{-3}$). 
For larger St, dust decouples from the gas, \added[id=R1]{settles to the disk midplane and accretes efficiently onto the planet}. 
This further produces dust depletion in the polar region of the Bondi sphere, but enhances the dust density at the midplane of the Bondi sphere (Fig.~\ref{fig:slice_xz_rhog_rhod_azimuthal_average}b and c; ${\rm St}=\,10^{-2}$ and $10^{-1}$).
\added[id=R3]{Although Fig.~\ref{fig:slice_xz_rhog_rhod_azimuthal_average}c shows a disk-like spatial distribution of dust, the azimuthal velocity of dust within the envelope remains strongly sub-Keplerian and smaller than the radial infall velocity (Fig.~\ref{fig:1dslice_vr_vphi}b).
Therefore, dust transport within the envelope is dominated by radial infall rather than rotational support.}

The radial motion of dust within the envelope shows no significant difference between the 2D and 3D runs (Fig. \ref{fig:1dslice_vr_vphi}). This similarity arises because the dust infall velocity \added[id=R1]{only depends on the planet mass or the Stokes number} (Eq. \ref{eq:v_d,in}).

Because of the vertical redistribution of dust, the dust-to-gas density ratio in 3D is higher at the midplane than in the purely 2D case, yielding a heterogeneous dust-to-gas ratio within the Bondi sphere with a maximum at the midplane. Figure \ref{fig:1dslice_d_to_g}b shows the dust-to-gas ratio \added[id=R1]{as a function of radius} in the 3D runs, \added[id=R1]{either shell averaged or azimuthally averaged at the midplane}. Nontheless, as in 2D, the ratio decreases toward the inner envelope. This is because vertical settling inherited from the gas dynamics enhances dust density by at most factor of 10, whereas the gas density increases exponentially inward. In Sect.~\ref{sec:1D analytic model}, we will introduce a detailed 1D model to reproduce the numerical result (dotted curve in Fig. \ref{fig:1dslice_d_to_g}b).

%---------------------------------------------------------
%---------------------------------------------------------
\subsection{\added[id=R1]{Simulations with fixed-size particles}}\label{sec:Simulations with fixed-size particles}

So far we have presented results assuming a fixed Stokes number. We now relax this assumption and consider fixed dust sizes. The dust stopping time is computed by Eq. \ref{eq:stopping time} and depends on the dust size, \added[id=R2]{the gas density,} and the mean free path of the gas. Consequently, for a given dust size, the local Stokes number varies within the envelope. 
To compute the dust stopping time \added[id=R4]{at the initial state}, \added[id=R5]{we adopt a passively irradiated disk model at 10 au \citep{oka2011evolution},
\begin{align}
    &\Sigma_0=2.3\times10^2\,\text{g/cm}^2\,\bigg(\frac{a_{\rm p}}{10\,\text{au}}\bigg)^{-15/14},\,T_0=56\,\text{K}\,\bigg(\frac{a_{\rm p}}{10\,\text{au}}\bigg)^{-3/7}.\label{eq:oka disk model}
\end{align}
Here we choose 10 au as a representative outer-disk location, where nearly isothermal envelopes are expected to occur \citepalias[Appendices A and B of][]{kuwahara2026multi2}.}
We then compute the sound speed and the gas density by \added[id=R1]{\added[id=R5]{$c_{\rm s,0}=\sqrt{k_{\rm B}T_0/(\mu m_{\rm H})}\simeq4.44\times10^4\,\mathrm{cm/s}$ and $\rho_{\rm g}=\Sigma_0/(\sqrt{2\pi}H_{\rm g,0})\simeq1.29\times10^{-11}\,\mathrm{g/cm^3}$}}, with $k_{\rm B}$ being the Boltzmann constant. 
We assumed 0.01, \added[id=R4]{0.1}, 1, and 10 cm-sized dust with $\rho_\bullet=3\,\text{g/cm}^3$, \added[id=R5]{corresponding to initial Stokes numbers of ${\rm St}=3.3\times10^{-4},\,3.3\times10^{-3},\,3.3\times10^{-2}$, and $0.33$, respectively.}

We find no significant differences between the fixed-St and the fixed-size runs. Figure \ref{fig:1dslice_fixed_size} summarizes the numerical results obtained from the 2D simulations. 
\added[id=R4]{The Stokes number varies within the envelope.}
The infall velocity of dust responds to variations in the Stokes number (Fig.~\ref{fig:1dslice_fixed_size}b).
A reduction in the infall velocity of dust leads to an increase in the local dust density and hence the dust-to-gas ratio relative to the fixed-St runs \added[id=R5]{(Figs.~\ref{fig:1dslice_d_to_g}a and \ref{fig:1dslice_fixed_size}c).}

\added[id=R2]{We note that, for $s=0.01$\,cm dust, the simulation does not reach a steady state within the calculation time.
We find that a small, but nonzero radial gas motion near the inner boundary affects the dust dynamics when $s\lesssim0.01$ cm \added[id=R5]{(corresponding to ${\rm St}\lesssim10^{-4}$)}.
Within our fiducial setup, the results for $s\geq0.1$\,cm are therefore physically robust, \added[id=R5]{and thus we omit the $s=0.01$\,cm case from Fig.~\ref{fig:1dslice_fixed_size}}.
The numerical difficulties associated with such small dust grains are discussed in Appendix~\ref{sec:Limitations for simulations with very small dust grains}.
}

Based on these fixed-size results and their close agreement with the fixed-St behavior, we proceed in the next section to develop a 1D model for dust dynamics in the envelope under the \added[id=R2]{terminal or free-fall velocity} approximation, which proves to be a match to the numerical results.

%---------------------------------------------------------
%---------------------------------------------------------
\section{1D (semi-)analytic models for gas and dust}\label{sec:1D analytic model}

Motivated by the 2D and 3D simulations in Sects. \ref{sec:Dynamics of gas and dust in 2D}--\ref{sec:Simulations with fixed-size particles}, here we construct a 1D model of gas and dust within an envelope. 
Our goal is to derive (semi-)analytic expressions for the dust density and, in combination with the gas density, obtain a model for the dust-to-gas density ratio \added[id=R1]{as a function of the radius. All analytic formulae are written using the dimensionless units introduced in Sect.~\ref{sec:Code units, simulation parameters, and initial condition}, in which $H_{\rm g,0}=c_{\rm s,0}=\Omega_0=\rho_{\rm g,0}=1$ and $\rho_{{\rm d},0}=0.01\,\rho_{\rm g,0}$.}

\added[id=R2]{Assuming azimuthal symmetry and ignoring the Coriolis and tidal terms, in 2D cylindrical coordinate, vortensity conservation and force balance give \citep{Ormel:2015a}:
\begin{align}
    &\frac{\partial (rv_{{\rm g},\phi})}{\partial r}=r\bigg(\frac{\rho_{\rm g}^{\rm 2D}}{2}-2\bigg),\label{eq:vortensity conservation}\\
    &\frac{1}{\rho_{\rm g}^{\rm 2D}}\frac{\partial \rho_{\rm g}^{\rm 2D}}{\partial r}=\frac{v_{{\rm g},\phi}^2}{r}-\frac{\partial \Phi_{\rm p}}{\partial r}.\label{eq:radial force balance}
\end{align}
This system holds for $r\lesssim R_{\rm B}$, where circular motion dominates the gas flow field. Solving these equations numerically with the boundary conditions, $\rho_{\rm g}^{\rm 2D}(R_{\rm B})=\rho_{\rm g,0}$ and $v_{{\rm g},\phi}(r_{\rm in})=0$ yields $\rho_{\rm g}^{\rm 2D}\!(r)$. In 3D, assuming the hydrostatic equilibrium, the gas density follows the isothermal limit \added[id=R4]{(Fig.~\ref{fig:1dslice_rho_and_temp})},
\begin{align}
    \rho_{\rm g}^{\rm 3D}(r)\approx\rho_{\rm g,0}\exp\Bigg(\frac{R_{\rm B}}{\sqrt{r^2+r_{\rm sm}^2}}\Bigg)\equiv\rho_{\rm g,iso}(r).\label{eq:gas density model, 3D}
\end{align}
}

\begin{figure}[tp]
    \centering
    \includegraphics[width=1\linewidth]{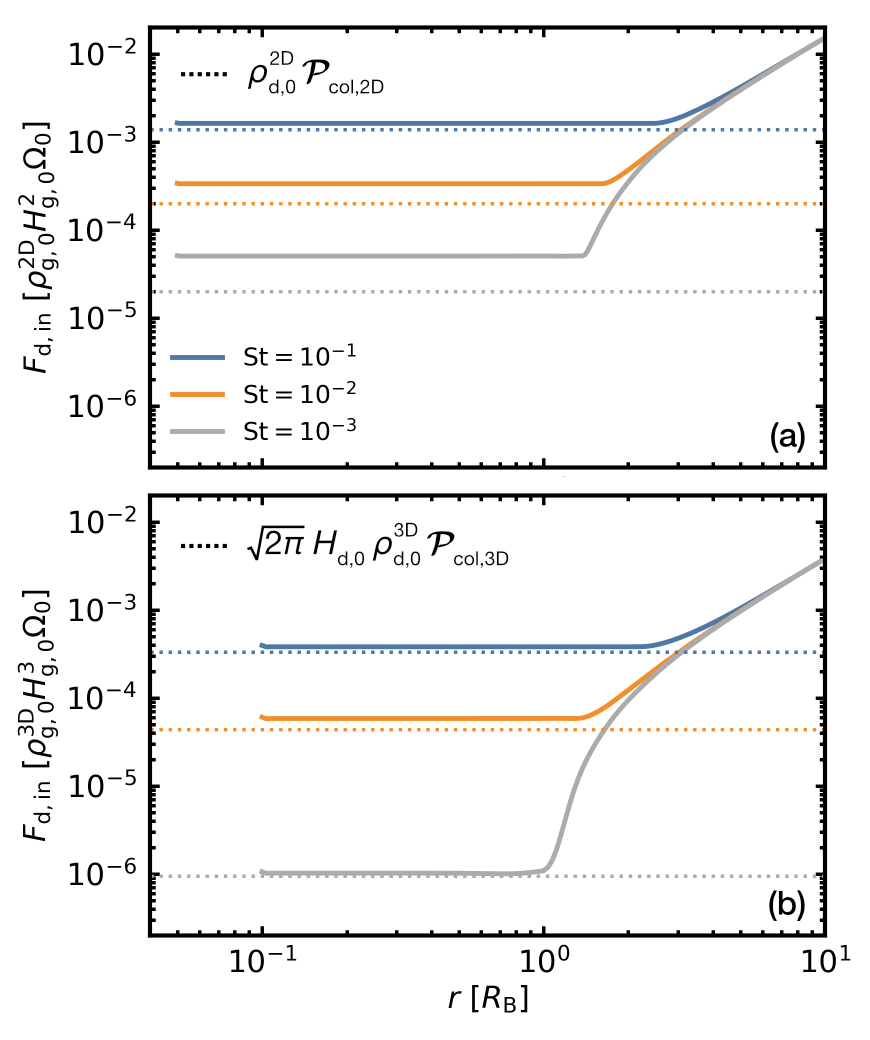}
    \caption{Inward dust mass flux obtained from the 2D (\textit{top}, azimuthally averaged) and 3D (\textit{bottom}, shell averaged) runs. The vertical axis is normalized by the value at the outer edge of the computational domain. \added[id=R4]{The dotted curves are given by Eqs.~\ref{eq:F_d,in 2Db} (top) and \ref{eq:F_d,in 3Db} (bottom), representing the accretion rate of incoming dust.}
    }
    \label{fig:1dslice_dust_flux}
\end{figure}

We assume that the inward dust mass flux, $F_{\rm d,in}$, is radially constant within an envelope. Here $F_{\rm d,in}$ is given by\footnote{We only considered the upper hemisphere in 3D. To obtain the numerically computed dust mass flux in 3D, we multiplied 2 assuming symmetry.}
\begin{empheq}[left = {F_{\rm d,in}=\empheqlbrace \,}]{alignat = 2}
    &2\pi rv_{\rm d, in}(r)\rho_{\rm d}^{\rm 2D}\!(r)&&\quad{\text{in the 2D cylindrical}}\label{eq:F_d,in 2Da},\\
    &4\pi r^2v_{\rm d, in}(r)\rho_{\rm d}^{\rm 3D}\!(r)&&\quad{\text{in the 3D spherical polar}}.\label{eq:F_d,in 3Da}
\end{empheq}
The solid curve in Fig. \ref{fig:1dslice_dust_flux} is the numerically calculated dust mass flux, confirming an $r$-independent dust mass flux within the envelope, $r\lesssim R_{\rm B}$. 
Outside the envelope, the inward mass flux of dust converges to the shear-dominated value.

\added[id=R4]{The dust accretion rate onto the core can be expressed as
\begin{empheq}[left = {\dot{M}_{\rm d,acc}=\empheqlbrace \,}]{alignat = 2}
    &\rho_{\rm d,0}^{\rm 2D}\,\mathcal{P}_{\rm col,2D}\left({\rm St}|_{r=R_{\rm B}}\right),\label{eq:F_d,in 2Db}\\
    &\sqrt{2\pi}H_{\rm d,0}\rho_{\rm d,0}^{\rm 3D}\,\mathcal{P}_{\rm col,3D}\left({\rm St}|_{r=R_{\rm B}},H_{\rm d,0}\right).\label{eq:F_d,in 3Db}
\end{empheq}
Here $\mathcal{P}_{\rm col}$ denotes the specific collision rate of dust grains, incorporating the effects of the planet-induced, non-Keplerian gas flow \citep[][see Appendix~\ref{sec:collision rate of pebbles} for the detailed expressions]{okamura2021growth}. }
\added[id=R4]{In the 3D case, we assumed that the unperturbed surface density is given by $\sqrt{2\pi}H_{\rm d,0}\rho_{\rm d,0}^{\rm 3D}$ (Eq.~\ref{eq:density profile}).}

\added[id=R4]{For fixed-size dust grains, both the stopping time and the terminal velocity vary within the envelope. 
The inward dust velocity is taken to be $v_{\rm d,in}(r)=\min(v_{\rm term},v_{\rm ff})$, with $v_{\rm ff}$ given by Eq.~\ref{eq:v ff}. 
\added[id=R5]{In the nonlinear regime, $t_{\rm s}$ depends on the relative speed.
The terminal velocity is then defined implicitly by
\begin{align}
    v_{\rm term}^2\,C_{\rm D}(v_{\rm term})
    =
    \frac{8}{3}\frac{\rho_\bullet s}{\rho_{\rm g}(r)}\,\frac{GM_{\rm p}}{r^2}.\label{eq:implicit relation}
\end{align}
At each radius we solve Eq.~\ref{eq:implicit relation} numerically by root finding.}}

Equating \added[id=R4]{\Equref{eq:F_d,in 2Da} and \Equref{eq:F_d,in 2Db} (or equivalently \Equref{eq:F_d,in 3Da} and \Equref{eq:F_d,in 3Db} in 3D)} yields the dust density within the envelope,
\added[id=R4]{\begin{align}
    &\rho_{\rm d}^{\rm 2D}\!(r)=\frac{\rho_{\rm d,0}^{\rm 2D}\,\mathcal{P}_{\rm col,2D}\left({\rm St}|_{r=R_{\rm B}}\right)}{2\pi rv_{\rm d, in}(r)},\label{eq:rhod model 2d}\\
    &\rho_{\rm d}^{\rm 3D}\!(r)=\frac{\sqrt{2\pi}H_{\rm d,0}\rho_{\rm d,0}^{\rm 3D}\,\mathcal{P}_{\rm col,3D}\left({\rm St}|_{r=R_{\rm B}},H_{\rm d,0}\right)}{4\pi r^2v_{\rm d, in}(r)}.\label{eq:rhod model 3d}
\end{align}}\added[id=R2]{For fixed-size dust grains, we evaluate the \added[id=R4]{collision rate} with the corresponding Stokes number at $r=R_{\rm B}$.}

The dust-to-gas density ratio within the envelope is computed with the set of analytic formulae for $\mathcal{P}_{\rm col},\,v_{\rm d, in}(r),\,\rho_{\rm g}^{\rm 3D}\!(r),\,\rho_{\rm d}^{\rm 2D}\!(r)$, and $\rho_{\rm d}^{\rm 3D}\!(r)$. We note that $\rho_{\rm g}^{\rm 2D}(r)$ must be computed numerically from vortensity conservation \added[id=R4]{together with radial force balance, and also $v_{\rm d, in}(r)$ if the nonlinear drag regime is considered}. 
\added[id=R4]{
The dotted curves in Fig.~\ref{fig:1dslice_d_to_g} show the 1D analytic model, which agrees with the numerical results of the fixed-St runs in both 2D and 3D, \added[id=R5]{as well as fixed-size runs.}}

Although the model presented in this study assumes a nearly isothermal, convectively stable envelope, it can be extended to adiabatic, convective envelopes by adopting appropriate expressions for the gas density and the \added[id=R2]{collision rate} \added[id=R3]{(Appendix~\ref{sec:collision rate of pebbles})}.

%---------------------------------------------------------
\section{Comparison to previous studies}\label{sec:Comparison to previous studies}
A key feature of our study is that we explicitly include the planet-perturbed gas flow when evaluating the dust mass flux inside the envelope. Previous studies often assume unperturbed \added[id=R1]{Keplerian} disk gas, which can overestimate the dust flux into the envelope. 
\added[id=R2]{\citet{Popovas:2018a} performed local 3D gas and dust simulations using super-particles.
Although their primary goal was to measure the accretion rate of particles onto the planet, they found no significant accumulation of particles inside the convectively stable envelope:
small grains are advected by the planet-induced gas flow, while larger grains settle efficiently onto the planet.
These results are consistent with our findings.}

Our simulations are currently limited to convectively stable envelopes and do not yet model convection.
Convection is triggered when the temperature gradient exceeds the adiabatic gradient \citep{rafikov2006atmospheres,piso2014minimum}, a condition favored in regions where small grains are abundant. Collisions and erosion are efficient processes that can generate a large amount of tiny grains ($<0.01$ cm) within the envelope \citep{ali2020limits,brouwers2021planets}. Because these small grains settle slowly, the dust opacity can rise, potentially rendering the envelope fully convective if dust accretion rates are sufficiently high \citep[$>10^{-5}\,M_\oplus/$yr][]{brouwers2021planets}. However, dust dynamics inherited from gas dynamics were not included in these studies. Such \added[id=R1]{an efficient supply of small dust grains} would be difficult, because they tend to follow the gas without entering the envelope (Fig. \ref{fig:slice_xy_d_to_g_2d}).

\cite{johansen2020transport} investigated transport of dust in a fully radiative envelope with a 1D model taking into account dust physics (dust growth, erosion, and fragmentation), finding that the dust-to-gas ratio remains nearly constant over a wide radial range.
This is because, in the Epstein drag regime, the Stokes number decreases inward as the gas density increases, and settling becomes inefficient.
%Our fixed-size runs capture this trend (gray and black solid curves in Fig. \ref{fig:1dslice_fixed_size}c). 
\added[id=R1]{\cite{johansen2020transport} considered a non-isothermal envelope (with an $r$-dependent sound speed), which further reduces the settling efficiency.}
If dust size reduction due to erosion were included in our models, the resulting dust-to-gas ratio would likely be higher than those shown in Fig.~\ref{fig:1dslice_fixed_size}.
\cite{johansen2020transport} also investigated dust transport in fully convective envelopes, finding that dust settling is inhibited and the dust-to-gas ratio increases radially inward. \added[id=R1]{The characteristic speed of convection reaches approximately 1--10\% of the sound speed at $r=R_{\rm B}$ \citep{KL26}, and thus would affect the infalling dust grains when $v_{\rm term}(R_{\rm B})\lesssim0.01\text{--}0.1c_{\rm s,0}$ (${\rm St}\lesssim10^{-3}\text{--}10^{-2}$; Eq.~\ref{eq:v_d,in}).} Multifluid simulations in such convective envelopes will be included in the second paper of this series \added[id=R3]{\citepalias{kuwahara2026multi2}}.

\cite{krapp20223d} performed global multifluid simulations for sub-thermal ($m=0.6$) and super-thermal ($m=1.8$) mass planets, finding that a strong latitudinal gradient in the dust distribution within the Hill sphere. 
This is caused by the density wave and polar inflow of the gas induced by the planet. 
The dust-to-gas ratio peaks at the disk midplane and is minimal at the pole, consistent with our 3D results (Fig. \ref{fig:slice_xz_rhog_rhod_azimuthal_average}). 
We note that our local simulations cannot fully capture the density waves, which extend over the full azimuthal range of a disk. 
A further methodological difference is that \added[id=R2]{we allow for dust transport across the inner envelope simulation boundary, in this way removing dust grains from the simulation domain, while this is not done in \cite{krapp20223d}.}
Therefore, radially-inward dust depletion inside the envelope is not observed in their simulations.

\added[id=R3]{
More massive gap-opening planets develop rotationally-supported envelopes, that can lead to the formation of sub-Keplerian circumplanetary disks \citep{lambrechts2019quasi}. 
This latter process has been shown to be promoted when effective cooling times are within orbital timescales \citep{krapp2024thermodynamic}. 
However, in our 3D simulations with a fixed cooling time, around lower-mass embedded planets, we find that the rotational gas velocity remains subdominant (Fig.~\ref{fig:1dslice_vr_vphi}b). 
Possibly, closer to the central core, in the interior that is not resolved in this study ($\lesssim 0.05\,R_{\rm B}$),  a disk-like structure could develop under near-isothermal conditions \citep{Fung:2019}.
This motivates the continued need for high-resolution studies with a self-consistent coupling between dust distribution and cooling times.
}

\added[id=R3]{\citet{takaoka2023spin} post-processed 3D hydrodynamical simulations to integrate pebble trajectories within a convectively stable envelope, finding trajectories consistent with ours (Fig.~\ref{fig:slice_xy_d_to_g_2d}).
They further showed that pebbles efficiently transfer prograde spin angular momentum to the core, as they are dragged by the prograde rotation of the envelope.
Although we do not compute the spin angular momentum explicitly, the prograde azimuthal dust velocities (Fig.~\ref{fig:1dslice_vr_vphi}) support the idea that pebble accretion in convectively stable envelopes naturally favors prograde planetary rotation \citep{Johansen:2010,visser2020spinning,yzer2023influence}.
}

\added[id=R4]{
Finally, we examine the dependence on the cooling time. 
In our fiducial 3D runs, the boundary between the recycling and convectively stable layers is located near the Bondi radius (the recycling–radiative boundary; RRB; Fig.~\ref{fig:slice_rhog_d_to_g_3d}a). 
The RRB depends on the cooling time \citep{Kurokawa:2018,bailey2024growing,KL26}. 
\citet{KL26} proposed the fitting formula
\begin{align}
    &r_{\rm RRB}^{\rm fit}=\min\Bigg(R_{\rm atm}^{\rm KK24},\,R_{\rm atm}^{\rm KK24}\times\beta^{0.22}\Bigg)\quad(\beta\leq1),\\
    &R_{\rm atm}^{\rm KK24}=C_1\,R_{\rm B}\Bigg(1-\frac{D_1}{C_1}\frac{\eta v_{\rm K}/c_{\rm s}}{R_{\rm B}/H}\Bigg).
\end{align}
Here $R_{\rm atm}^{\rm KK24}$ is the fitting formula for the size of the radiative envelope \citep{kuwahara2024analytic}, where $C_1=0.84$ and $D_1=0.056$ are the fitting coefficients, $\eta$ is a dimensionless quantity characterizing the global pressure gradient of the disk gas, and $v_{\rm K}$ is the Keplerian speed.
The polar inflow is unable to penetrate beyond $r_{\rm RRB}^{\rm fit}$, because a positive entropy gradient (buoyancy force) suppresses the inflow penetrating the envelope \citep{Kurokawa:2018}.
As $\beta$ decreases, the RRB shifts inward. 
In the isothermal limit, where buoyancy is absent, the recycling flow can penetrate deeper into the envelope.
In our fiducial 3D runs with $\beta=1$, the RRB lies near $R_{\rm B}$, and the radius where the dust-to-gas ratio sharply decreases also coincides with $R_{\rm B}$ (Fig.~\ref{fig:1dslice_d_to_g}b). 
This transition radius is expected to move inward as $\beta$ decreases, as is confirmed in Appendix~\ref{sec:Additional simulations}.
}

%---------------------------------------------------------
%---------------------------------------------------------
\section{Discussions} \label{sec:Discussions}

%---------------------------------------------------------
\subsection{Implications for thermal evolution of envelopes}\label{sec:Implications for thermal evolution of envelopes}
\added[id=R3]{An envelope around an embedded planet cools, contracts, and continually accretes gas, eventually reaching the point of runaway gas accretion} \citep[e.g.,][]{Mizuno:1980,Pollack:1996}. 
The corresponding timescale, $t_{\rm run}$, is sensitive to several parameters, including the core mass and the dust opacity 
\citep[e.g.,][]{stevenson1982formation,ikoma2000formation}:
\begin{align}
    t_{\rm run}\approx 3\times10^5\,\mathrm{yr}\,\Bigg(\frac{M_{\rm p}}{10\,M_\oplus}\Bigg)^{-2.5}\Bigg(\frac{\kappa_{\rm d}}{1\,\mathrm{cm^2/g}}\Bigg).
\end{align}
\added[id=R3]{Therefore,} a reduction in dust opacity shortens $t_{\rm run}$ \citep{Hori:2011,Lee:2014,ormel2021planets}.
Dust is a major source of the opacity and contributes in proportion to the dust-to-gas ratio,
\begin{align}
    \kappa_{\rm d}=\frac{3Q}{4\rho_\bullet s}\frac{\rho_{\rm d}}{\rho_{\rm g}}.
\end{align}
Here $Q=\min(0.6\pi s/\lambda_{\rm max},\,2)$ is the extinction efficiency with $\lambda_{\rm max}$ being the peak wavelength from Wien's displacement law. 
The dust opacity in the outer envelope inherits the background disk value, which can be assumed to be the ISM-like value of approximately 1 cm$^2$/g \added[id=R3]{\citep{bell1994using}.}

\added[id=R3]{O}ur results show that, in a convectively stable envelope, the dust opacity decreases radially inward by orders of magnitude relative to this outer value. 
\added[id=R3]{Moreover,} \added[id=R1]{the anisotropic distribution of dust shown in Fig.~\ref{fig:slice_xz_rhog_rhod_azimuthal_average} allows polar radiation escape, which may further enhance envelope cooling \citep{krapp2024thermodynamic}.} 
\added[id=R2]{A full analysis of the effective cooling rate of the planet would require explicit radiative transfer modeling, taking into account the anisotropic distribution of dust, with significant polar depletion (Fig.~\ref{fig:slice_xz_rhog_rhod_azimuthal_average}). 
Such vertical radiation escape may further enhance envelope cooling \citep{krapp2024thermodynamic}.}
\added[id=R2]{Because} radiative layers are expected to develop preferentially in the outer disk \citep[$\gtrsim10$ au][]{KL26}, dust depletion within radiative envelopes can facilitate the early onset of runaway gas accretion, potentially aiding the formation of outer gas giants. 
Moreover, additional physics not included in this study such as dust sublimation and the associated envelope enrichment further shorten $t_{\rm run}$ \citep{Lambrechts:2014,venturini2016planet}.

\added[id=R1]{Finally, in this work we enforced a convectively stable envelope by adopting a finite cooling time $\beta$ independent of the dust content. 
Our results imply that, once a radiative layer is established, dust is efficiently depleted from it, which further shortens the cooling time by lowering the opacity. 
Consequently, unless a substantial amount of small grains is supplied to the envelope, the radiative layer is likely to persist.}

%---------------------------------------------------------
\subsection{Dust processing within envelopes}\label{sec:Dust processing within envelopes}
Throughout this study we considered fixed-St or fixed-size dust and neglected dust physics such as growth, fragmentation, ablation, sublimation, and erosion. \added[id=R1]{Dust depletion is the generic outcome in a convectively stable envelope, primarily because large dust settles efficiently and the envelope is shielded from the incoming flux of small grains.} Dust growth within an envelope enhances dust settling, leading to further dust depletion \citep{ormel2014atmospheric,mordasini2014grain}.

\added[id=R1]{However, as discussed in Sect.~\ref{sec:Comparison to previous studies}, dust depletion may be inhibited under particular circumstances.
If large pebbles accrete at high rates and fragment or erode into micron–submicron grains, the resulting tiny particles have extremely long settling times and can raise the local dust-to-gas ratio \citep{ali2020limits, brouwers2021planets}.}
\added[id=R5]{Indeed, the terminal-velocity approximation predicts settling times of order $10^3\,\Omega_0^{-1}$ for 1 micron-sized grains in the reference disk model at 10 au (Eq.~\ref{eq:oka disk model}), substantially longer than the duration of our simulations.
The hard-to-determine abundance of such grains likely depends on the evolutionary history of the planet and envelope.
Dust settling may be more efficient during the early stages of planet growth, when the Bondi radius is smaller, whereas recycling flows can filter newly arriving grains at later stages.
Micron-sized grains may also be replenished by dust processing within the envelope.
A self-consistent prediction of their abundance would therefore require following the coupled evolution of the planet, gas, and dust during planetary growth.}

\added[id=R1]{Ablation and sublimation of species during settling is not included in this study. However, we expect that these thermal processes have a limited impact on our results. Outside the water snowline, the water sublimation front lies deep inside the envelope \citep[$<0.1\,R_{\rm B}$;][]{wang2023atmospheric}. Silicates sublimates even closer to the core surface \citep{brouwers2020planets}. Thermal processing therefore becomes important only at radii well inside our computational domain \citep{Alibert:2017,brouwers2018cores,valletta2019deposition,vazan2023rocky,lous2024accretion}.}

%---------------------------------------------------------
\subsection{Implications for atmospheric metallicity in planets}\label{sec:Implications for atmospheric metallicity in planets}
Thanks to the James Webb Space Telescope (JWST), we can now obtain detailed measurements of the atmospheric metallicity of exoplanets, defined as the abundance of elements heavier than helium. 
Recent JWST observations have revealed several exoplanets exhibiting sub-stellar atmospheric metallicities, characterized by O/H ratios lower than those of their host stars \citep[][and compiled in \citealp{ohno2026dichotomy}]{taylor2023awesome,fournier2024near,fournier2025transmission,smith2024combined,meech2025bowie,davenport2025toi,liu2025unveiling}. 

Such sub-stellar atmospheric metallicities can naturally arise if planets accrete gas located beyond the water snowline, where oxygen is largely sequestered in icy grains, leaving the disk gas oxygen-poor \citep[e.g.,][]{schneider2021drifting,bitsch2022drifting,danti2023composition}. 
A pebble accretion-based population synthesis model further predicts that planets forming beyond the water snowline can retain sub-stellar atmospheric metallicities if their envelopes remain poorly mixed owing to inefficient convection \citep{ohno2026dichotomy}. 
Our results lend additional support to this interpretation. The envelopes around embedded planets at distant locations are likely to be convectively stable \citep{KL26}. 
The outer regions of such envelopes \added[id=R1]{($>0.1\,R_{\rm B}$)} are expected to be dust- \added[id=R1]{and volatile-}poor \added[id=R3]{during their formation.}

\added[id=R2]{
Nevertheless, our study only focuses on the disk-embedded stage of planet formation.
Assessing whether primordial compositions are preserved in observed exoplanet atmospheres requires accounting for post-disk evolutionary processes, including long-term mixing within the envelope \citep{honing2019carbon,schlichting2022chemical,vazan2024planets,werlen2025atmospheric,steinmeyer2026coupled}.}

%---------------------------------------------------------
%---------------------------------------------------------
\section{Conclusions} \label{sec:Conclusions}
We have conducted a suite of two- and three-dimensional multifluid (gas and dust) simulations to characterize the spatial distribution of \added[id=R3]{solids} within the nearly isothermal, convectively-stable envelope around \added[id=R2]{an} Earth-like planet embedded in a disk. 
Our high-resolution simulations resolve both the disk gas flow perturbed by the planet and the interior structure of the envelope, including an inner, convectively stable layer that is largely shielded from recycling flows.

We identify a dust-depleted envelope as a robust outcome of our models.
The dust-to-gas ratio decreases monotonically toward the inner envelope: at $0.1\,R_{\rm B}$, the local dust-to-gas ratio is reduced by more than two to four orders of magnitude relative to its value at $R_{\rm B}$.
This trend is insensitive to the assumed dust Stokes number (or size).
On the one hand, large grains with ${\rm St}\gtrsim10^{-2}$ enter the envelope, settle efficiently toward the core, and are rapidly accreted.
On the other hand, small grains with ${\rm St}\lesssim10^{-3}$ remain tightly coupled to the disk gas flow and largely avoid entering the envelope.
As a result, the envelope becomes depleted of both small and large dust grains relative to the gas.
We further constructed 1D analytic models for the dust-to-gas ratio within the envelope, which successfully reproduce the numerical results.

These findings have direct implications for the envelope’s thermal evolution.
Because dust depletion strongly lowers the opacity, the envelope cools efficiently, potentially accelerating the onset of runaway gas accretion.
\added[id=R2]{At the same time our work argues that \added[id=R3]{volatile enrichment of the inner envelope} requires midplane-accreting pebbles that sublimate, or catastrophically fragment, inside of the outer convectively stable layer ($\lesssim0.1\,R_{\rm B}$).
This opens a pathway for volatile-depleted outer atmospheres, provided no efficient envelope mixing occurs after disk dissipation.}

To conclude, our work highlights the need to follow coevolution of gas and dust for a deeper understanding of planet-envelope systems.
In this study, we fixed the envelope cooling time to keep the envelope nearly isothermal and convectively stable.
\added[id=R5]{These idealized models isolate the role of envelope dynamics in dust transport by prescribing the cooling time.}
In reality, the cooling time is set by the opacity, whose primary source is small dust grains, while the dust distribution itself is controlled by the envelope-scale gas flow.
A \added[id=R3]{future} self-consistent treatment therefore requires coupling envelope gas dynamics, dust transport, and thermodynamics.

%-------------------------------------------------------------------------------------------
\begin{acknowledgements}
\added[id=R5]{We thank the anonymous referee for an exceptionally thorough and insightful review, whose detailed comments substantially improved the quality of this manuscript.}
We thank the Athena++ developers. The Tycho supercomputer hosted at the SCIENCE HPC center at the University of Copenhagen was used for supporting this work. M.L. \added[id=R2]{acknowledges} the ERC starting grant 101041466-EXODOSS. \added[id=R2]{We are grateful for helpful} discussions with Ziyan Xu and Kazumasa Ohno.
\end{acknowledgements}
%

%-------------------------------------------------------------------------------------------

%% references
%%\raggedright              %% only for adsaa with dvips, not for pdflatex
%\input{accretion_heating.bbl} 
%\ref{Ref}
\bibliography{Ref_aanda}

\clearpage
\begin{appendix}
%\appendix
%\setcounter{section}{1}
\def\thesection{A}
\setcounter{equation}{0}
\def\theequation{A.\arabic{equation}}
\setcounter{figure}{0}
\def\thefigure{A.\arabic{figure}}

\section{Convergence tests}\label{sec:Convergence tests}
\begin{figure}[tp]
    \centering
    \includegraphics[width=1\linewidth]{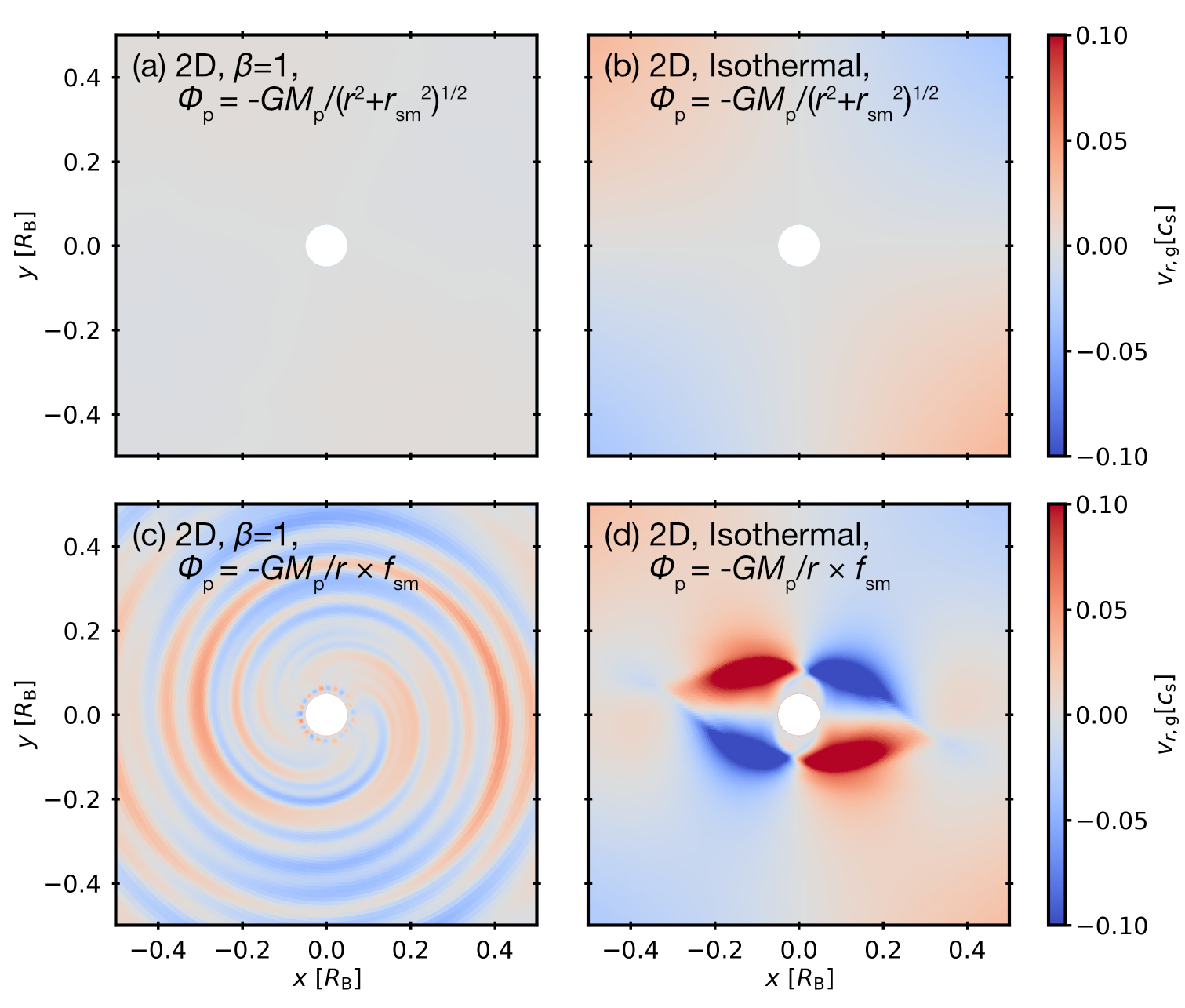}
    \caption{Radial gas velocity in the deep envelope for different smoothing prescriptions under \added[id=R1]{nearly} isothermal ($\beta=1$) and isothermal conditions. Panel a shows the fiducial setup.}
    \label{fig:convergence_test_grav_potential}
\end{figure}

\begin{figure}[tp]
    \centering
    \includegraphics[width=1\linewidth]{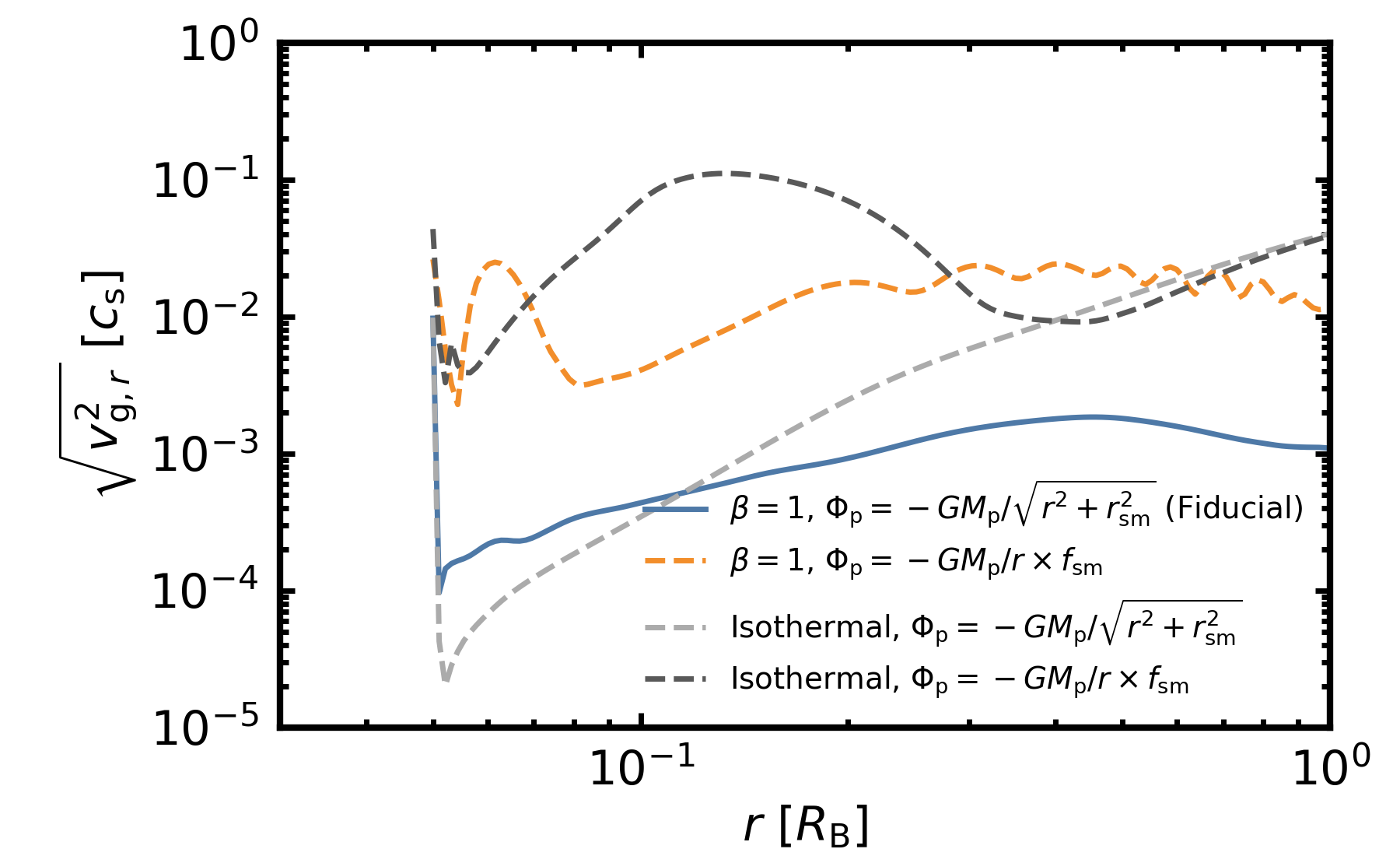}
    \caption{\added[id=R2]{Root-mean-squared radial gas velocity as a function of radius}, obtained from the same data shown in Fig. \ref{fig:convergence_test_grav_potential}.}
    \label{fig:convergence_test_vr}
\end{figure}

%-------------------------------------------------------------------------------------------
In this section, we investigate how the simulation results depend on numerical settings: the gas isothermality, the resolution, the size of the inner boundary, the gravitational smoothing, and the inclusion of the tidal force. Table \ref{tab:hydro simulations} summarizes the parameter choices used for the convergence tests.

We find that the gas isothermality and the gravitational smoothing have a crucial impact on the numerical results. Throughout this study, we focus on the nearly isothermal, convectively stable envelopes. Ideally, \added[id=R1]{hydrostatic equilibrium is established inside the envelope, in which the radial gas velocity becomes negligible.}

Figure \ref{fig:convergence_test_grav_potential} shows the radial gas velocity in the deep envelope and compares 2D fiducial runs with isothermal runs for different smoothing prescriptions (Plummer and force-free at $r_{\rm in}$). 
Only our fiducial model (Fig. \ref{fig:convergence_test_grav_potential}a; $\beta=1$ with Plummer smoothing) exhibits a negligible radial gas velocity, whereas in all other cases a prominent radial gas motion is observed within the envelope. 
\added[id=R2]{
This radial gas motion, possibly caused by the numerical boundary effects, can be seen in previous study and would disappear in a simulation with a larger smoothing length or a higher-resolution \citep{Ormel:2015b}.}
This spurious $v_{r,{\rm g}}$ affects dust dynamics and can lead to unphysical dust accumulation. 
We note that, however, even in the fiducial run $v_{r,{\rm g}}$ is nonzero.
\added[id=R2]{Figure~\ref{fig:convergence_test_vr} shows the root-mean-squared of $v_{r,{\rm g}}$ inside the envelope, of which intensity is on the order of $\sim10^{-3}\,c_{\rm s,0}$ in the fiducial run.
We will discuss the effect of this small, but nonzero $v_{r,{\rm g}}$ on the dust motion in Appendix~\ref{sec:Limitations for simulations with very small dust grains}.
}

\begin{figure}[tp]
    \centering
    \includegraphics[width=1\linewidth]{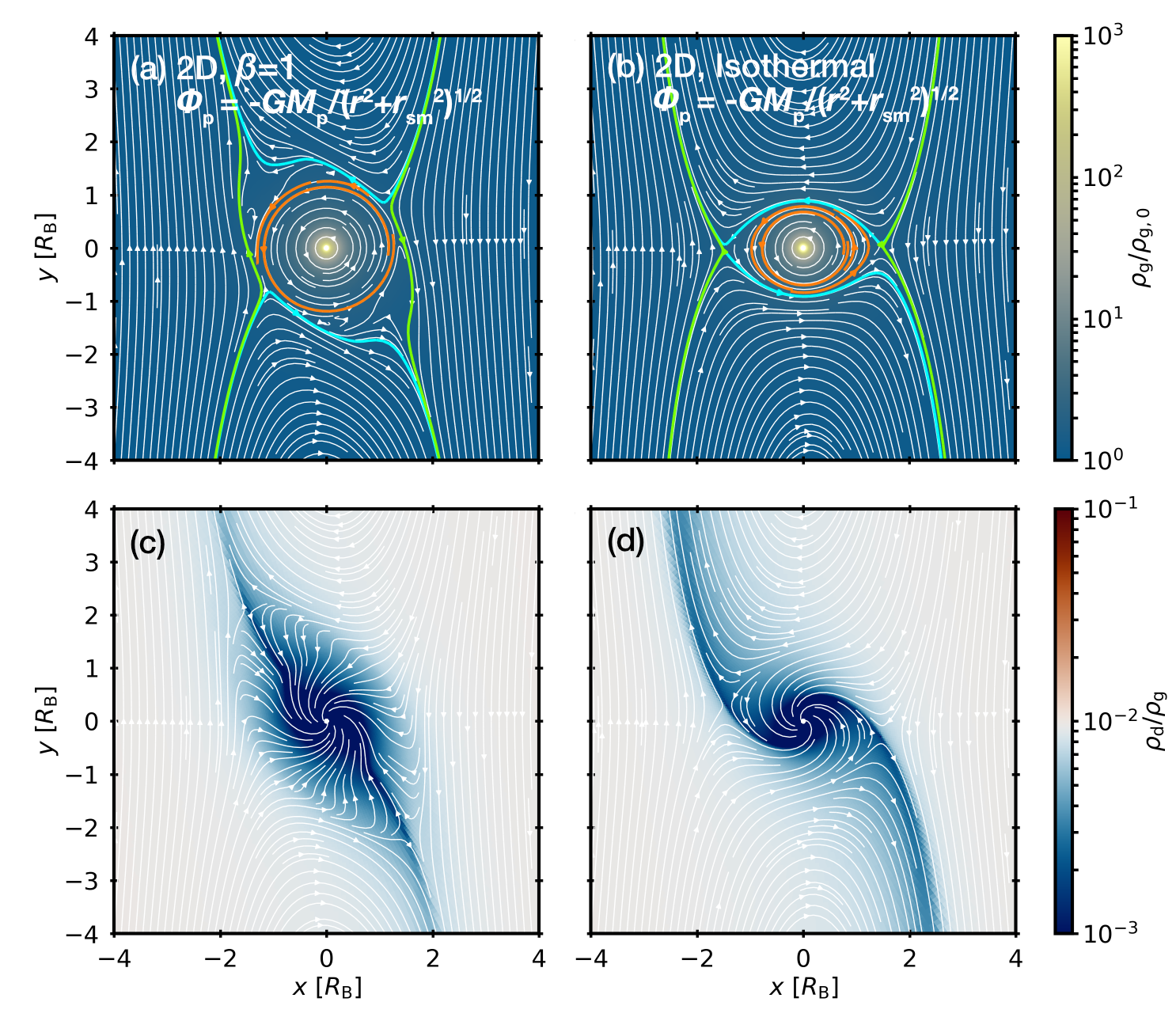}
    \caption{Flow fields of gas and dust around an embedded planet in the 2D runs under \added[id=R1]{nearly} isothermal ($\beta=1$) and isothermal conditions. The left column corresponds to the fiducial setup. We adopt the Plummer smoothing. \textit{Top}: Gas surface density with gas streamlines. \added[id=R1]{The orange, cyan, and green curves mark the outer envelope, outer horseshoe and inner shear streamlines, respectively.} \textit{Bottom}: Column dust-to-gas ratio with dust streamlines (${\rm St}=10^{-2}$).}
    \label{fig:convergence_test_iso_beta}
\end{figure}

In 2D, adopting a finite $\beta$ value alters the streamline topology relative to purely isothermal runs. Figures \ref{fig:convergence_test_iso_beta}a and b compare midplane streamlines for the fiducial and isothermal cases. With $\beta=1$, the envelope characterized by closed gas streamlines is larger, and the horseshoe region is non-axisymmetric. These differences in gas flow lead to different dust dynamics, and consequently to variations in the extent of the dust-depleted region. Neverthless, the azimuthally averaged dust-to-gas ratio agrees within at most a factor of 2 between fiducial and isothermal runs (Fig. \ref{fig:convergence_test_d_to_g}).

\begin{figure}[tp]
    \centering
    \includegraphics[width=1\linewidth]{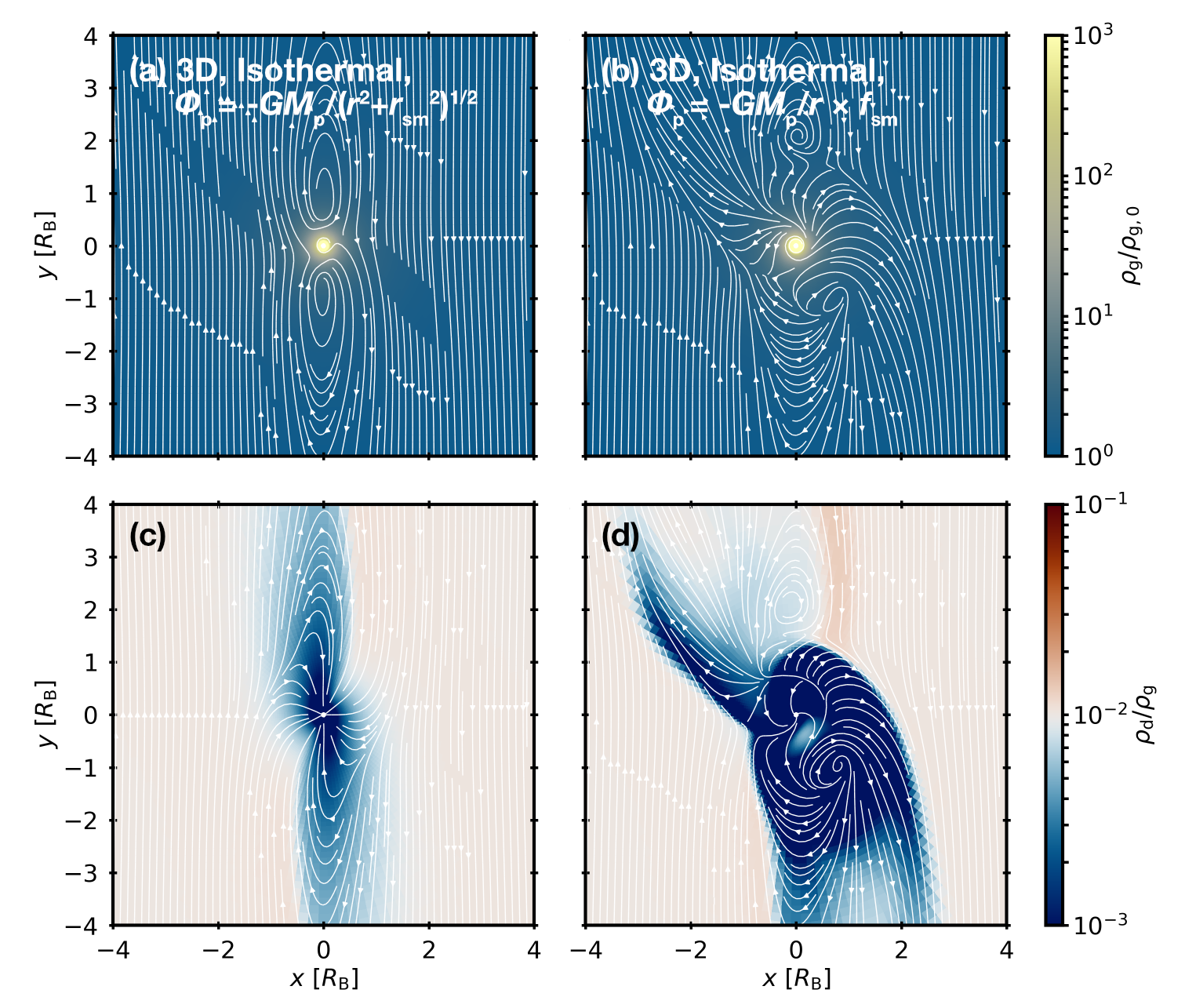}
    \caption{Midplane slices of the gas and dust flow field in the 3D runs for different smoothing prescriptions under isothermal conditions. These setups are not used in the main text. \textit{Top}: Gas density with gas streamlines. \textit{Bottom}: Dust-to-gas ratio with dust streamlines (${\rm St}=10^{-2}$).}
    \label{fig:convergence_test_3d}
\end{figure}

\begin{figure}[tp]
    \centering
    \includegraphics[width=1\linewidth]{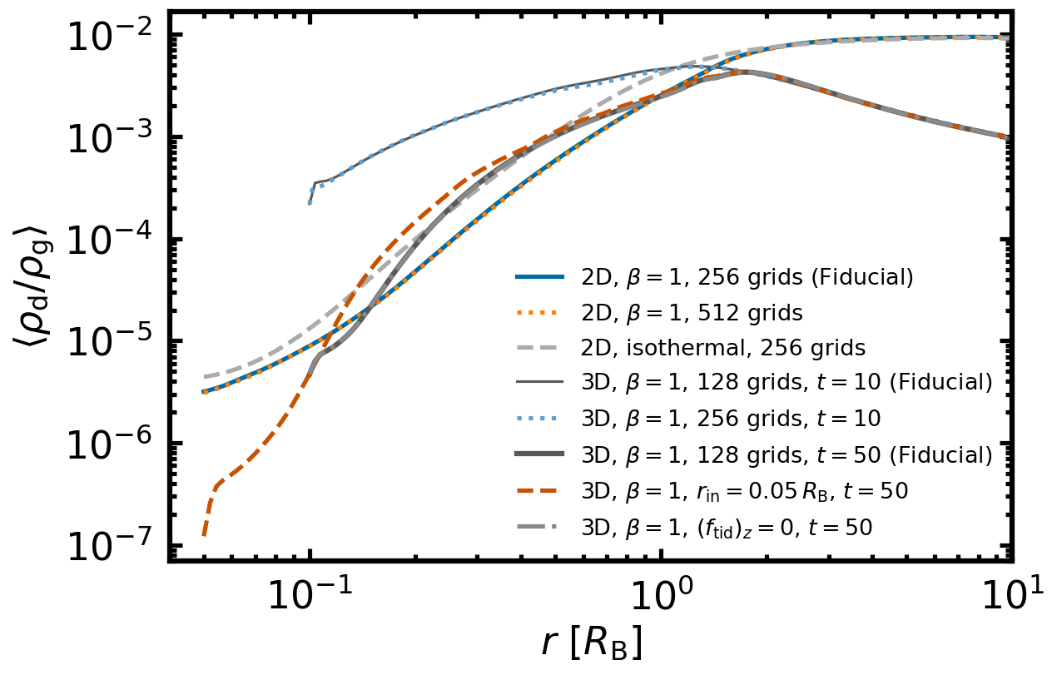}
    \caption{Azimuthally and shell averaged dust-to-gas ratio in the 2D and 3D runs for different numerical settings. This figure compares our fiducial setup (solid curves) with the high-resolution runs (dotted), the 2D isothermal run (gray dashed), the run with a smaller $r_{\rm in}$ (brown dashed), and the run without the vertical component of the tidal force (dot-dashed). We set ${\rm St}=10^{-2}$.}
    \label{fig:convergence_test_d_to_g}
\end{figure}

In 3D, \added[id=R1]{pure} isothermal runs with Plummer smoothing reach a steady state but develop vortices in the horseshoe region (Fig. \ref{fig:convergence_test_3d}a), consistent with a previous work \citep{Kuwahara:2019}. With force-free smoothing at $r_{\rm in}$, the gas flow field does not reach a steady state under isothermal condition (Fig. \ref{fig:convergence_test_3d}b). These gas dynamics influence the dust dynamics (Figs. \ref{fig:convergence_test_3d}c and d). The dependence on the gravitational smoothing has been extensively examined in previous studies, which concluded that employing the force-free smoothing is the most appropriate choice \citep{Fung:2019,zhu2021global,KL26}. In our numerical tests, the gas flow field reaches a quasi-steady state only when a finite $\beta$ is adopted together with the force-free potential at $r_{\rm in}$.

Physically, isothermal envelopes lack a buoyancy barrier against polar gas inflow because of the absence of an entropy gradient. Consequently, the polar inflow penetrates deep into the envelope, and there is no clear boundary demarcating the bound atmosphere from the disk gas. Although the cooling time depends on opacity, non-isothermal conditions are more realistic. With finite $\beta$, a buoyancy barrier isolates the inner envelope, making comparison with 2D results more straightforward. 

For these reasons, our fiducial setup uses $\beta=1$ with Plummer (force-free at $r_{\rm in}$) smoothing in 2D (3D). Figure \ref{fig:convergence_test_d_to_g} presents additional convergence tests for the fiducial setup (solid curves), showing the dependence on the resolution and the inner boundary size. The fiducial results are numerically converged and insensitive to $r_{\rm in}$ within the tested range.

Finally, we note that, in post-processing calculations that integrate pebble trajectories in a given gas flow field without turbulent stirring, the vertical component of the tidal force, which enhances dust settling, is often neglected  \citep{Visser:2016,Kuwahara:2020a,takaoka2023spin}. Although our multifluid simulations differ from these post-processed simulations, neither include turbulent diffusion, making the comparison straightforward. We confirmed that the inclusion or omission of the vertical component of the tidal force does not significantly affect our results (gray dot-dashed curve in Fig. \ref{fig:convergence_test_d_to_g}). This is because our computational domain size of $r_{\rm out}=10\,R_{\rm B}$ is much smaller than the aforementioned studies, typically they assume $40\,R_{\rm H}\simeq12.8\,r_{\rm out}$, and therefore the dust settling due to the vertical tidal force is inefficient in our simulations.

%-------------------------------------------------------------------------------------------
\def\thesection{B}
\setcounter{equation}{0}
\def\theequation{B.\arabic{equation}}
\setcounter{figure}{0}
\def\thefigure{B.\arabic{figure}}

\section{Limitations for simulations with very small dust grains}\label{sec:Limitations for simulations with very small dust grains}
\begin{figure}[tp]
    \centering
    \includegraphics[width=1\linewidth]{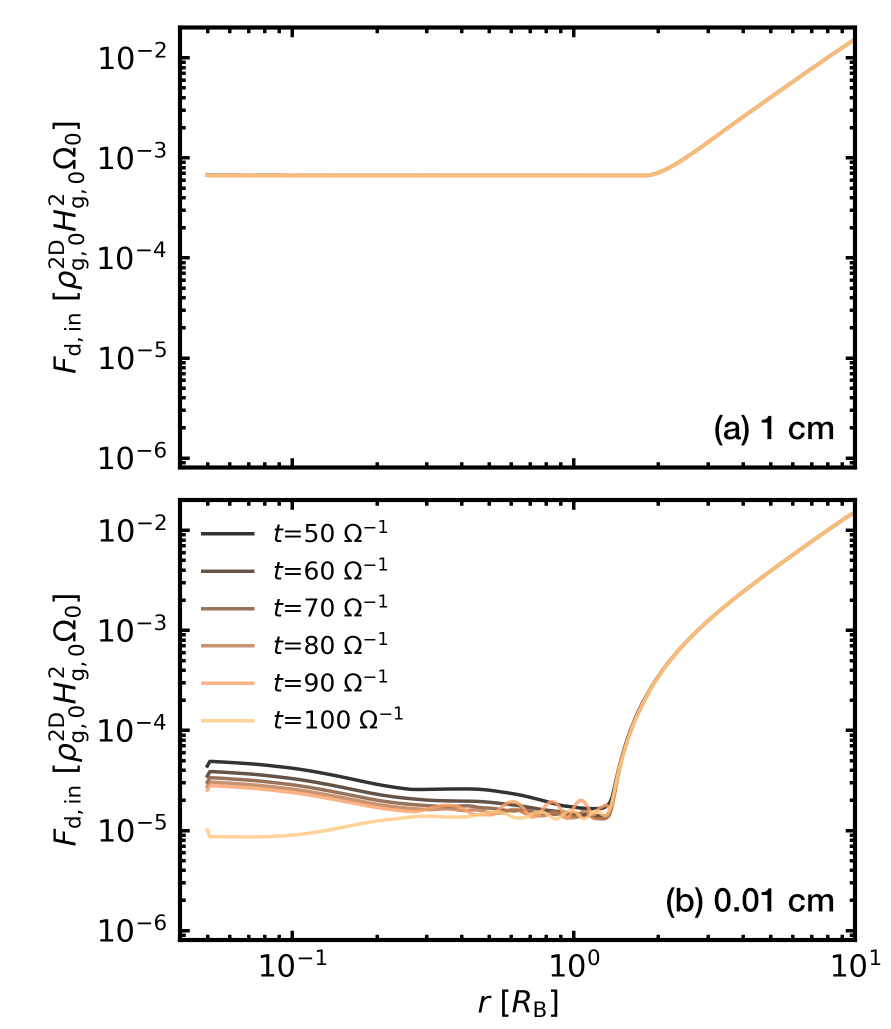}
    \caption{Azimuthally averaged radial dust mass flux obtained from the 2D, fixed dust size run.}
    \label{fig:dust_massflux_appendix}
\end{figure}

We confirm that all fiducial runs reach a steady state, except for the run with the smallest dust size, $s=0.01$\,cm (Fig.~\ref{fig:dust_massflux_appendix}).
In this section, we discuss numerical limitations that arise when simulating dust with very small Stokes numbers.

When the Stokes number is extremely small, dust dynamics becomes highly sensitive to the gas velocity field inside the envelope.
Even after the envelope gas has reached hydrostatic equilibrium, a nonzero radial gas velocity component of order $\sim10^{-3}\,c_{\rm s,0}$ remains and cannot be fully eliminated with our fiducial numerical setup (Fig.~\ref{fig:convergence_test_vr}).
If the terminal velocity of dust is smaller than this residual gas velocity, dust dynamics becomes dominated by numerical effects.
This situation occurs for very small grains.
\added[id=R5]{We indeed confirmed that the numerically obtained infall velocity of $s=0.01$\,cm-sized dust deviates from the terminal velocity prediction (not shown in Fig.~\ref{fig:1dslice_fixed_size}b).}
Therefore, within the fiducial numerical setup, our results are physically robust for dust sizes $s\gtrsim0.1$\,cm, corresponding to ${\rm St}\gtrsim10^{-4}$, whereas simulations with smaller grains are increasingly affected by numerical artifacts.

%-------------------------------------------------------------------------------------------
\def\thesection{C}
\setcounter{equation}{0}
\def\theequation{C.\arabic{equation}}
\setcounter{figure}{0}
\def\thefigure{C.\arabic{figure}}

\begin{figure}[tp]
    \centering
    \includegraphics[width=1\linewidth]{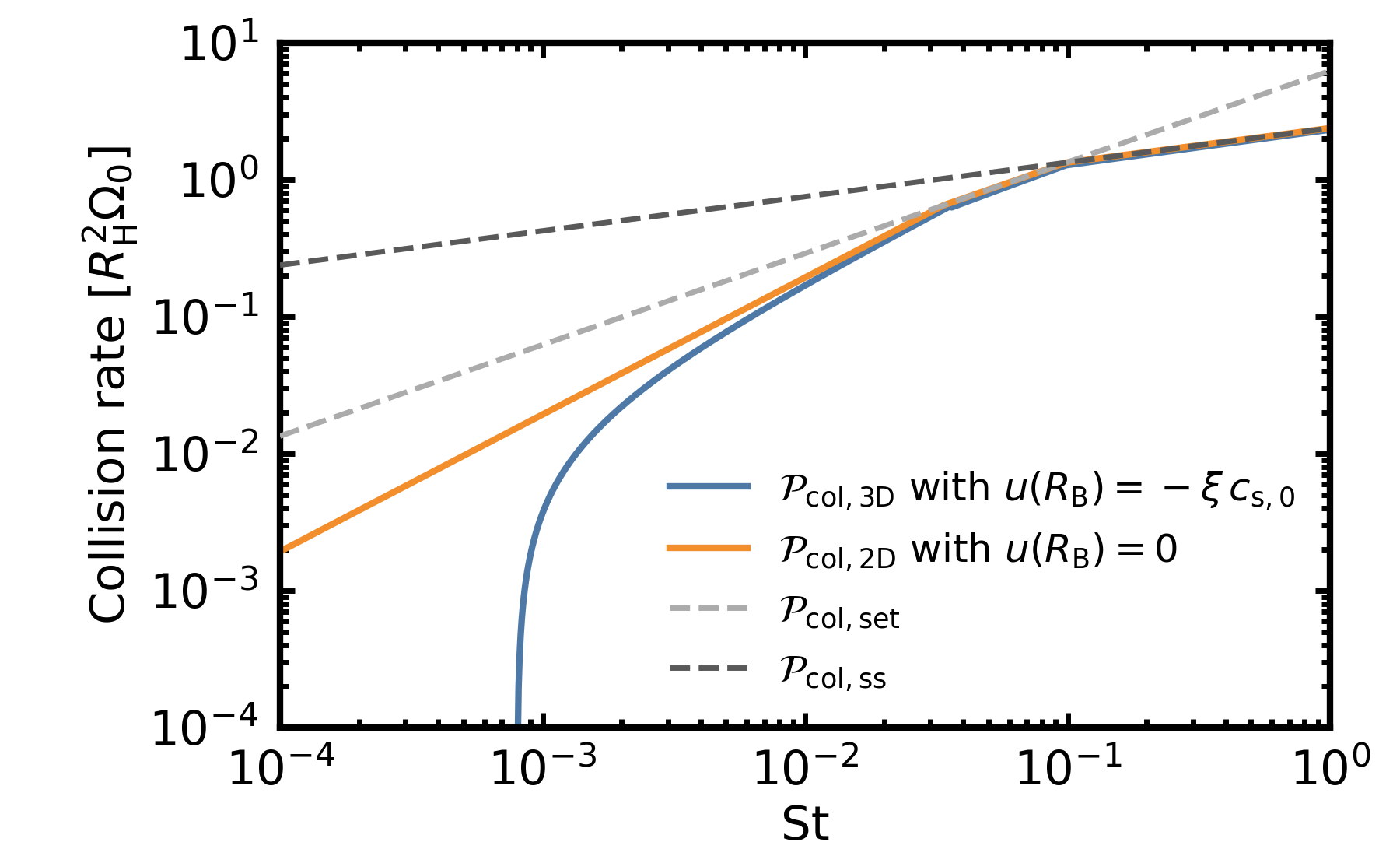}
    \caption{\added[id=R4]{Specific collision rate of pebbles computed from Eq.~\ref{eq:pcol appendix}. The recycling flow associated with the midplane outflow suppresses the accretion of small pebbles with ${\rm St}\lesssim10^{-3}$ (blue curve).}}
    \label{fig:pcol_p1}
\end{figure}

\section{\added[id=R4]{Specific collision rate of pebbles}}\label{sec:collision rate of pebbles}
\added[id=R2]{Here we briefly summarize the \added[id=R4]{specific collision rate} of pebbles in a non-Keplerian gas flow perturbed by a planet. Following \cite{okamura2021growth}, the \added[id=R4]{specific collision rate} is written as}
\begin{empheq}[left = {\mathcal{P}_{\rm col}=\empheqlbrace \,}]{alignat = 2}
    &\mathcal{P}_{\rm col,2D}({\rm St}),\nonumber\\
    &\mathcal{P}_{\rm col,3D}({\rm St},H_{\rm d,0}),\label{eq:pcol appendix}
\end{empheq}
and
\tiny
\added[id=R4]{
\begin{empheq}[left = {\empheqlbrace \,}]{alignat = 2}
    &\mathcal{P}_{\rm col,2D}({\rm St})=\min\big(\mathcal{P}_{\rm col,set},\,\mathcal{P}_{\rm col,ho},\,\mathcal{P}_{\rm col,ss}\big),\\
    &\mathcal{P}_{\rm col,3D}({\rm St},H_{\rm d,0})=\Bigg[\bigg(\mathcal{P}_{\rm col,2D}({\rm St})\bigg)^{-2}+\Bigg(\mathcal{P}_{\rm col,2D}({\rm St})\frac{x_{\rm ss}}{0.65\,H_{\rm d,0}}\Bigg)^{-2}\Bigg]^{-1/2},\\
    &\mathcal{P}_{\rm col,set}=3\,(2C_1{\rm St})^{2/3}\,R_{\rm H}^2\Omega_0,\\
    &\mathcal{P}_{\rm col,ho}=2\frac{R_{\rm B}}{R_{\rm H}}\Bigg[\frac{3\,{\rm St}}{(R_{\rm B}/R_{\rm H})^2} + \frac{u(R_{\rm B})}{c_{\rm s,0}}\sqrt{\frac{3}{(R_{\rm B}/R_{\rm H})}}\Bigg]\,R_{\rm H}^2\Omega_0,\\
    &\mathcal{P}_{\rm col,ss}=2\sqrt{6}C_1\Bigg(\frac{\rho_{\bullet}}{\rho_{\rm g}}\frac{c_{\rm s}}{R_{\rm H}\Omega_0}\frac{l_{\rm mfp}}{R_{\rm H}}{\rm St}\Bigg)^{1/4}\,R_{\rm H}^2\Omega_0,\\
    &x_{\rm ss}=
    \begin{cases}
        (2C_1{\rm St})^{1/3}\,R_{\rm H}&\text{if } \mathcal{P}_{\rm col,2D}=\mathcal{P}_{\rm col,set},\\
        2\,R_{\rm B}&\text{if } \mathcal{P}_{\rm col,2D}=\mathcal{P}_{\rm col,ho},\\
        \displaystyle
        (8/3)^{1/4}\sqrt{C_1}\Bigg(\frac{\rho_{\bullet}}{\rho_{\rm g}}\frac{c_{\rm s}}{R_{\rm H}\Omega_0}\frac{l_{\rm mfp}}{R_{\rm H}}{\rm St}\Bigg)^{1/8}\,R_{\rm H}&\text{if } \mathcal{P}_{\rm col,2D}=\mathcal{P}_{\rm col,ss}.
    \end{cases}
\end{empheq}}\normalsize with $C_1=1.5$.
Here, $u(R_{\rm B})$ parameterizes the characteristic radial gas motion at the Bondi radius \citep{okamura2021growth},
\begin{align}
    u(R_{\rm B})=-\xi\,c_{\rm s,0},
\end{align}
with
\added[id=R4]{\begin{empheq}[left ={\xi=\empheqlbrace \,}]{alignat = 2}
    &0\quad&&\text{in 2D},\\
    &0.08\,m\quad&&\text{in 3D}.
\end{empheq}}

\added[id=R2]{We slightly modified the original formula given by \cite{okamura2021growth}, where the authors set $
\xi=0.1\,m$. This parameter controls the influence of the midplane gas outflow onto accreting dust, reducing the \added[id=R4]{collision rate} \citep{Kuwahara:2019}. 
Because such midplane outflow associated with the 3D recycling flow is absent in 2D, we set $\xi=0$. 
In 3D, we adopt \added[id=R4]{$\xi=0.08\,m$}, chosen to match our numerical results.}
\added[id=R2]{We additionally ignore the \added[id=R4]{collision rate in the gas-free regime, relevant for ${\rm St}>10^{0}$ \citep[defined as $\mathcal{P}_{\rm atm}$ in][]{okamura2021growth}.}  
See Table 2 in \cite{okamura2021growth} for a complete formula.}\added[id=R4]{
The collision rate in the supersonic regime, $\mathcal{P}_{\rm col,ss}$, is evaluated using $\rho_{\rm g}$, $c_{\rm s}$, $\Omega_0$, and $l_{\rm mfp}$ at 1~au in the passively irradiated disk model adopted in Sect.~\ref{sec:Simulations with fixed-size particles}.
Because this term is only relevant for ${\rm St}\gtrsim m$ \citep{okamura2021growth}, the collision rate in our setup is primarily determined by $\mathcal{P}_{\rm col,set}$ or $\mathcal{P}_{\rm col,ho}$ (Fig.~\ref{fig:pcol_p1}).
}

Finally, we note that the \added[id=R4]{collision-rate} formalism summarized here can be further generalized to convectively unstable envelopes.
In particular, the effects of convective flows can be incorporated into the \added[id=R4]{collision rate} through an appropriate choice of the characteristic radial gas velocity $u(R_{\rm B})$.
\added[id=R3]{Such an extension will be introduced in the second paper of this series \citepalias{kuwahara2026multi2}.}

%-------------------------------------------------------------------------------------------
\def\thesection{D}
\setcounter{equation}{0}
\def\theequation{D.\arabic{equation}}
\setcounter{figure}{0}
\def\thefigure{D.\arabic{figure}}

\section{\added[id=R4]{Additional simulations}}\label{sec:Additional simulations}

\begin{figure}[tp]
    \centering
    \includegraphics[width=1\linewidth]{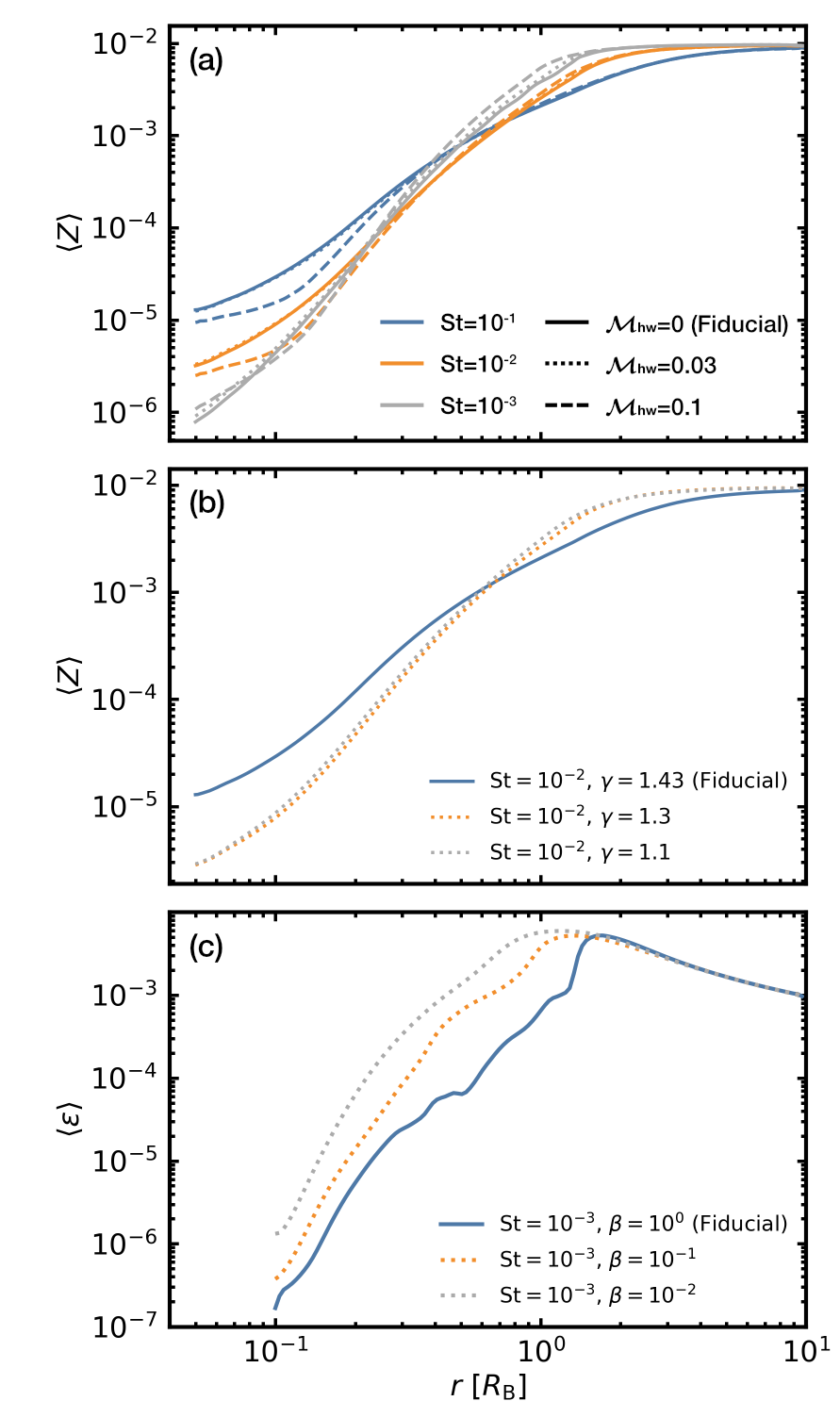}
    \caption{\added[id=R4]{Dust-to-gas ratio for different numerical setups. The top panel compares the fiducial run (solid) with headwind runs (dotted and dashed), the middle panel shows runs with different adiabatic indices (dotted), and the bottom panel shows the dependence on $\beta$.}}
    \label{fig:convergence_test_hw_gamma_beta}
\end{figure}

\begin{table*}[htbp]
\caption{Parameters of hydrodynamical simulations.}
%\centering
\resizebox{\textwidth}{!}{
\begin{tabular}{lccccccccc}\hline\hline
         & $m$ & St & $\beta$ & Resolution & $t_{\rm end}$ [$\Omega_0^{-1}$] & $r_{\rm in}$ [$R_{\rm B}$] & $r_{\rm out}$ [$R_{\rm B}$] & $\mathcal{M}_{\rm hw}$ & $\gamma$ \\ \hline
     Fiducial runs (2D)  & $0.1$ & $10^{-3},\,10^{-2},\,10^{-1}$ & $10^0$ & $(N_r,\,N_\phi)=(256,\,256)$ & 100 & $0.05$ & $10$ & 0 & 1.43 \\
     Fiducial runs (3D)  & $0.1$ & $10^{-3},\,10^{-2},\,10^{-1}$ & $10^0$ & $(N_r,\,N_\theta,\,N_\phi)=(128,\,32,\,128)$ & 100 & $0.1$ & $10$ & 0 & 1.43\\\hline
     Different headwind runs  & $0.1$ & $10^{-3},\,10^{-2},\,10^{-1}$ & $10^0$ & $(N_r,\,N_\phi)=(256,\,256)$ & 100 & $0.05$ & $10$ & 0.03, 0.1 & 1.43 \\
     Different $\gamma$ runs  & $0.1$ & $10^{-2}$ & $10^0$ & $(N_r,\,N_\phi)=(256,\,256)$ & 100 & $0.05$ & $10$ & 0 & 1.1, 1.3 \\
     Different $\beta$ runs  & $0.1$ & $10^{-3}$ & $10^{-2},\,10^{-1}$ & $(N_r,\,N_\theta,\,N_\phi)=(128,\,32,\,128)$ & 100 & $0.1$ & 10 & 0 & 1.43 \\\hline     
\end{tabular}
}
\tablefoot{\added[id=R4]{The following columns give the dimensionless thermal mass, the Stokes number, the dimensionless cooling time, the resolution, the calculation time, the size of the inner boundary, the size of the outer boundary, the Mach number of the headwind, and the adiabatic index.}}
\label{tab:additional hydro simulations}
\end{table*}

\added[id=R4]{Here we summarize additional simulations that include physical effects not considered in the fiducial runs (\Tabref{tab:additional hydro simulations}).
In realistic disks, the gas rotates at a sub-Keplerian speed owing to the global pressure gradient, which induces radial drift of dust.
This effect was neglected in the fiducial runs.}

\added[id=R4]{We included the headwind through an additional source term, $\bm{f}_{\rm src}=\bm{f}_{\rm grav}+\bm{f}_{\rm cor}+\bm{f}_{\rm tid}+\bm{f}_{\rm hw}$, where $\bm{f}_{\rm hw}=2\mathcal{M}_{\rm hw}\bm{e}_x$ represents the global pressure-gradient force and $\mathcal{M}_{\rm hw}$ is the headwind Mach number \citep[e.g.,][]{Kurokawa:2018}.
The initial velocities of gas and dust are modified as \citep{Weidenschilling:1977,Nakagawa:1986},
\begin{align}
    &\frac{\bm{v}_{\rm g,\infty}}{c_{\rm s,0}}=\Bigg(-\frac{3}{2}\frac{x}{H_{\rm g,0}}-\mathcal{M}_{\rm hw}\Bigg)\bm{e}_y,\\
    &\frac{\bm{v}_{\rm d,\infty}}{c_{\rm s,0}}=-\frac{2\mathcal{M}_{\rm hw}{\rm St}}{1+{\rm St}^2}\bm{e}_x-\Bigg(\frac{3}{2}\frac{x}{H_{\rm g,0}}+\frac{\mathcal{M}_{\rm hw}}{1+{\rm St}^2}\Bigg)\bm{e}_y,
\end{align}
For the passively irradiated disk model adopted in Sect.~\ref{sec:Simulations with fixed-size particles}, the headwind Mach number is estimated as \citep{Ida:2016}
\begin{align}
    \mathcal{M}_{\rm hw}\simeq0.03\Bigg(\frac{L_\ast}{L_\odot}\Bigg)^{1/7}\Bigg(\frac{M_\ast}{M_\odot}\Bigg)^{-4/7}\Bigg(\frac{r_{\rm p}}{1\,\mathrm{au}}\Bigg)^{2/7},
\end{align}
assuming a solar-mass star with solar luminosity.
We considered $\mathcal{M}_{\rm hw}=0.03$ and $0.1$, spanning a broad range of the disk.
}

\added[id=R4]{Although the headwind modifies the flow pattern around an embedded planet \citep{Ormel:2013,Kurokawa:2018} and makes the dust accretion paths asymmetric \citep{benitez2018torques,Kuwahara:2020b}, its impact on the dust-to-gas ratio within the envelope remains minor (Fig.~\ref{fig:convergence_test_hw_gamma_beta}a).
This is because the headwind has little effect on the radial infall of dust once it has entered the envelope.}

\added[id=R4]{We note, however, that our local setup likely overestimates the dust accretion rate in the presence of a headwind. 
The accretion rate is expected to decrease with increasing headwind velocity \citep{Liu:2018}.
When radial drift is efficient, pebble accretion becomes asymmetric, with particles supplied predominantly from outside the planet’s orbit \citep[e.g., Fig.~7 of][]{Kuwahara:2020b}.
In a global disk, some particles would bypass the planet or be deflected away before reaching its immediate vicinity.
In our local setup, by contrast, the limited computational domain effectively initializes particles near the planet, allowing them to be accreted before this large-scale drift is fully realized.
Our setup therefore allows accretion from both sides and does not capture this large-scale drift.
Consequently, the dust-to-gas ratios obtained in this study should be regarded as upper limits.
}

\added[id=R4]{We also performed simulations with lower values of $\gamma$, motivated by the fact that the effective adiabatic index can be reduced in height-integrated 2D models \citep{goldreich1986stability,ostriker1992near,Gammie:2001}.
Figure~\ref{fig:convergence_test_hw_gamma_beta}b shows the dust-to-gas ratio in 2D runs with $\gamma=1.1$, $1.3$, and $1.43$ (fiducial).
All runs show the similar trends, because the radial infall of dust is not strongly affected by the choice of $\gamma$.
}

\added[id=R4]{Finally, Fig.~\ref{fig:convergence_test_hw_gamma_beta}c shows the dependence on the cooling time, $\beta$.
The envelope remains convectively stable for $\beta \lesssim 10$ \citep{KL26}.
As discussed in Sect.~\ref{sec:Comparison to previous studies}, the boundary between the recycling and radiative layers (RRB) shifts inward as $\beta$ decreases.
Because the recycling flow prevents small grains from penetrating the envelope, the radius at which the dust-to-gas ratio begins to decline also moves inward with decreasing $\beta$.
}

%------------------------------------------------------------------------------------------
\end{appendix}
%------------------------------------------
\end{document}